\documentclass[aps,prc,twocolumn,nofootinbib,floatfix,showpacs,preprintnumbers,amsmath,amssymb]
{revtex4}
\usepackage{graphicx}
\usepackage{bm}

\begin{document}
%
%
%
%
\newcommand{\covdev}{{\widetilde \partial}}
\newcommand{\finalnewpage}{\newpage}
\newcommand{\lagrang}{{\cal L}}
\newcommand{\newg}{{\skew2\overline g}_\rho}
\newcommand{\Nbar}{\skew3\overline N \mkern2mu}
\newcommand{\stroke}[1]{\mbox{{$#1$}{$\!\!\!\slash \,$}}}

\newcommand{\dgamma}[1]{\gamma_{#1}}
\newcommand{\ugamma}[1]{\gamma^{#1}}
\newcommand{\dsigma}[1]{\sigma_{#1}}
\newcommand{\usigma}[1]{\sigma^{#1}}
\newcommand{\ugammafive}{\gamma^{5}}
\newcommand{\dgammafive}{\gamma_{5}}
\newcommand{\ket}[1]{\vert#1\rangle}
\newcommand{\bra}[1]{\langle#1\vert}
\newcommand{\inprod}[2]{\langle#1\vert#2\rangle}
\newcommand{\psibar}[1]{\overline{#1}}
\newcommand{\half}{\ensuremath{\frac{1}{2}}}
\newcommand{\threehalf}{\ensuremath{\frac{3}{2}}}
\newcommand{\slashed}[1]{\not\!#1}
\newcommand{\lag}{\mathcal{L}}
\newcommand{\uhalftau}[1]{\frac{\tau^{#1}}{2}}
\newcommand{\dhalftau}[1]{\frac{\tau_{#1}}{2}}
\newcommand{\ucpartial}[1]{\widetilde{\partial}^{#1}}
\newcommand{\dcpartial}[1]{\widetilde{\partial}_{#1}}
\newcommand{\mn}{\mu\nu}
\newcommand{\ab}{\alpha\beta}
\newcommand{\Tdagger}[2]{T^{\dagger \,#1}_{#2}}
\newcommand{\T}[2]{T^{#1}_{\,\,#2}}
\newcommand{\vbg}{\mathsf{v}}
\newcommand{\abg}{\mathsf{a}}
\newcommand{\sbg}{\mathsf{s}}
\newcommand{\pbg}{\mathsf{p}}
\newcommand{\modular}[2]{\vert \vec{#1}_{#2}\vert}
\def\Tr{\mathop{\rm Tr}\nolimits}

%
%
\def\dthree#1{\intback{\rm d}^3\intback #1}
\def\dthreex{\dthree{x}}
\def\fpi{f_{\pi}}
\def\gammafive{\gamma^5}
\def\gammafivel{\gamma_5}
\def\gammamu{\gamma^{\mu}}
\def\gammamul{\gamma_{\mu}}
\def\intback{\kern-.1em}
\def\kfermi{k_{\sssize {\rm F}}}    
\def\mn{{\mu\nu}}
\def\sigmamunu{\sigma^\mn}
\def\sigmamunul{\sigma_\mn}
\def\Tr{\mathop{\rm Tr}\nolimits}

\let\dsize=\displaystyle
\let\tsize=\textstyle
\let\ssize=\scriptstyle
\let\sssize\scriptscriptstyle
%
%
%


\newcommand{\beq}{\begin{equation}}
\newcommand{\eeq}{\end{equation}}
\newcommand{\beqa}{\begin{eqnarray}}
\newcommand{\eeqa}{\end{eqnarray}}

\def\Inthelimit#1{\lower1.9ex\vbox{\hbox{$\
   \buildrel{\hbox{\Large \rightarrowfill}}\over{\scriptstyle{#1}}\ $}}}
\interfootnotelinepenalty=10000

\title{Neutrinoproduction of Photons and Pions From Nucleons in a Chiral 
            Effective Field Theory for Nuclei}

\author{Brian D. Serot} 
\author{Xilin Zhang}\email{xilzhang@indiana.edu}
\affiliation{Department of Physics and Center for Exploration of
             Energy and Matter\\
             Indiana University, Bloomington, IN\ \ 47405}

%
\author{\null}
\noaffiliation

%
\date{\today\\[20pt]}

\begin{abstract}
Neutrino-induced production (neutrinoproduction) of photons and pions  
from nucleons and nuclei
is important for the interpretation of neutrino-oscillation
experiments, as these photons and pions are potential backgrounds in the
MiniBooNE experiment [A. A. Aquilar-Arevalo \textit{et al.}
(MiniBooNE Collaboration), Phys.\ Rev.\ Lett.\ {\bf 100}, 032301
(2008)]. These processes are studied at intermediate
energies, where the $\Delta$ (1232) resonance 
becomes important. The
Lorentz-covariant effective field theory, which is the framework used in 
this series of studies, contains nucleons, pions,
$\Delta$s, isoscalar scalar ($\sigma$) and vector ($\omega$) fields,
and isovector vector ($\rho$) fields.  The Lagrangian exhibits a
nonlinear realization of (approximate) $SU(2)_L \otimes SU(2)_R$
chiral symmetry and incorporates vector meson dominance. 
In this paper, we focus on setting up the framework. 
Power counting for vertices and Feynman diagrams is
explained. Because of the built-in symmetries, the vector current
is automatically conserved, and the axial-vector current
is partially conserved. 
To calibrate the
axial-vector transition current $(N\! \leftrightarrow \Delta)$, pion
production from the nucleon is used as a benchmark and compared to
bubble-chamber data from Argonne and Brookhaven National
Laboratories. At low energies, the convergence of our power-counting
scheme is investigated, and next-to-leading-order tree-level
corrections are found to be small.
\end{abstract}

\smallskip
\pacs{25.30.Pt; 24.10.Jv; 11.30.Rd; 12.15.Ji}

\maketitle

\section{Introduction}
\label{sec:intro}

Neutrinoproduction of photons and pions from nucleons and nuclei plays an 
important role in the interpretation of neutrino-oscillation experiments, such
as MiniBooNE  \cite{MINIBOONE}. 
The neutral current (NC) $\pi^0$ and photon production produce detector signals 
that resemble those of the desired $e^\pm$ signals. 
Currently, it is still a question whether 
NC photon production might explain the excess events seen at low reconstructed neutrino energies 
in the MiniBooNE experiment, which the MicroBooNE experiment plans to answer \cite{MicroBN2011}.  
Moreover, pion absorption after production could lead to events 
that mimic quasielastic scattering. 

Ultimately, the calculations must be done on nuclei, which are the
primary detector materials in oscillation experiments. To separate
the many-body effects from the reaction mechanism and to calibrate
the elementary amplitude, we study charged current (CC) and NC
pion production from free nucleons in this work, which serves as the 
benchmark. Moreover, NC photon
production, which is not a topic under intense investigation, is
studied within this calibrated framework. In future papers, we
will include the electroweak response of the nuclear many-body
system to discuss the productions from nuclei in the same 
framework.

Here we use a recently proposed Lorentz-covariant
meson--baryon effective field theory (EFT) that was originally
motivated by the nuclear many-body problem
\cite{SW86,SW97,FST97,FSp00,FSL00,EvRev00,LNP641,EMQHD07}. (This
formalism is often called \emph{quantum hadrodynamics} or QHD.) This
QHD EFT includes all the relevant symmetries of the underlying QCD;
in particular, the approximate, spontaneously broken $SU(2)_L
\otimes SU(2)_R$ chiral symmetry is realized nonlinearly. The
motivation for this EFT and some calculated results are discussed in
Refs.~\cite{SW97,FST97,HUERTAS02,HUERTASwk,HUERTAS04,MCINTIRE04,
MCINTIRE05,JDW04,MCINTIRE07,HU07,MCINTIRE08,BDS10}.
In this EFT, we have the $\Delta$ resonance consistently incorporated 
as an explicit degree of freedom, while respecting the
underlying symmetries of QCD noted earlier. (The generation of mesons and
the $\Delta$ resonance through pion-pion interactions and
pion-nucleon interactions has been investigated in \cite{Lin89,
Lin91}.) We are concerned with
the intermediate-energy region $(E^{\mathrm{Lab}}_\nu \leqslant 0.5
\,\mathrm{GeV})$, where the resonant behavior of the $\Delta$
becomes important. The details about introducing 
$\Delta$ degree of freedom, the full Lagrangian, and electroweak interactions in this model  
have been presented in \cite{SZbookchapter,XZThesis}.
The well-known pathologies associated with introducing $\Delta$ are 
not relevant in the context of EFT.
The couplings to electroweak fields are included
using the external field technique \cite{Gasser84}, which allows us
to deduce the electroweak currents. Because of the approximate
symmetries in the Lagrangian, the vector currents are automatically conserved 
and the axial-vector currents are partially conserved.
Form factors are generated within the theory by vector meson
dominance (VMD), which allows us to avoid introducing phenomenological form
factors and makes current conservation manifest.\footnote{Meson
dominance generates form factors for contact pion-production 
vertices automatically, as shown in diagram (f) in 
Fig.~\ref{fig:feynmanpionproduction}. In other approaches, for example
\cite{HERNANDEZ07}, these form factors are introduced by
hand, which requires specific relations between the nucleon vector
current and the pion vector current form factors. This is explained
in Secs.~\ref{sec:currentmatrixelement} and
\ref{sec:diag}.} We discuss the power counting of both
vertices and diagrams on and off resonance and consistently keep all
tree-level diagrams through next-to-leading order. 
Explicit power-counting of loop diagrams in this EFT has
been discussed in Refs.~\cite{MCINTIRE07,HU07,MCINTIRE08}.
Here the contributions of the loops are assumed to be 
(mostly) saturated by heavy mesons and
the $\Delta$ resonance, so the couplings of contact interactions are viewed
as being renormalized. The mesons'
role in effective field theory has also been investigated in
\cite{ecker89_npb, ecker89_plb}. 

One major goal of this work is to calibrate electroweak interactions 
on the nucleon level. It is typically assumed that the vector part of the $N \to \Delta$
transition current is well constrained by electromagnetic
interactions \cite{OL06,GRACZYK08}. The uncertainty is in the
axial-vector part of the current, which is determined by fitting to Argonne National Laboratory (ANL) \cite{RADECKY82} and Brookhaven National
Laboratory (BNL) \cite{KITAGAKI86} bubble-chamber data.
The data have large error bars, which leads to significant model
dependence in the fitted results
\cite{HERNANDEZ07,GRACZYK09,PRAET09}.  Here we choose one
recently fitted parametrization \cite{GRACZYK09} and use it to
determine the constants of our VMD parametrization 
(but note that our basis of currents is different from the conventional 
one as used in \cite{HERNANDEZ07,GRACZYK09,PRAET09}). 
In addition, we make use of other form factors, 
the ones in \cite{HERNANDEZ07} for example. 
We then compare results of using different current basis and form factors with the data 
at low and intermediate neutrino energies.

There have been numerous earlier studies of neutrinoproduction of pions from
nucleons in the resonance region 
\cite{ADLER68,LS72,SCHREINER73,REIN81,AR99,SATO03,OL05,OL06,%
HERNANDEZ07,GRACZYK08,GRACZYK09,RHill09,OL10}. They basically fall
into two categories. In the first one \cite{REIN81, AR99,
OL05, OL06, GRACZYK08, GRACZYK09} resonance dominance above
intermediate energy is assumed. The contributions of resonances are summed 
incoherently and hence it is difficult to
determine the interference effect. In the second category, \cite{ADLER68,
SATO03,HERNANDEZ07,RHill09,OL10}, the contributions are summed 
coherently including the background,
since either an effective Hamiltonian or Lagrangian is utilized.

Our approach belongs to the second category, while differences from
other models should be mentioned. First, there
exists a finite energy range in which EFT is
valid, so we insist on low-energy calculations. However, a different
attitude has been taken, for example, in Refs.~\cite{HERNANDEZ07,
RHill09}, in which the Born approximation based on an effective
Lagrangian has been extrapolated to the region of several GeV.
Second, we have discussed the consequence of higher-order contact
terms.\footnote{Some of these terms have also been discussed in
Ref.~\cite{RHill09}; however, the interpretation of these
terms is different here from that in \cite{RHill09}.} Naturally,
these contributions should obey naive power counting
\cite{GEORGI84,GEORGI93}; however, some of them may play an important
role in scattering from nuclei. Third, electroweak 
interactions of nucleons are calibrated in this work while the 
strong interaction has been calibrated to nuclear properties. 
This is a unique feature that is absent in other models targeting 
the production from free nucleons only. Furthermore, the 
calibration on the nucleon's electroweak interaction 
impacts the strong interaction. For example,
the $\rho \, \pi \, \pi$ coupling, introduced because of VMD
in the pion's vector current, 
gives rise to an interesting contribution in the two-body
axial current in a  many-body calculation \cite{AXC02}.
In our theory
with $\Delta$, it can be quite interesting to investigate similar
consequences, for example, the $\Delta$'s role in the two-body
current, in which meson-dominance couplings can give rise to
relevant interactions.

This article is organized as follows: in Sec.~\ref{sec:LnoDelta} and~\ref{sec:LDelta}, we
introduce our Lagrangian without and with $\Delta$, 
and we calculate several current matrix
elements that will be useful for the subsequent Feynman diagram
calculations. The theory involving $\Delta$ is emphasized. 
Then the transition current basis and form factors are discussed carefully.
In Sec.~\ref{sec:diag}, we discuss our calculations for the CC and
NC pion production and for the NC photon production.
After that, we show our results in
Sec.~\ref{sec:res}. Whenever possible, we compare our results with
available data and present our analysis. Finally, our conclusions
are summarized in Sec.~\ref{sec:sum}.

In the Appendixes, we present the necessary information about chiral symmetry 
and electroweak interactions in QHD EFT, form factor calculations, power
counting for the diagram with $\Delta$, and kinematics.

\section{Lagrangian without $\Delta$ (1232)} \label{sec:LnoDelta}
In this work, the metric $g_{\mn}=\mathrm{diag}(1,-1,-1,-1)_{\mn}$. 
The convention for the Levi-Civita symbol $\epsilon^{\mu\nu\alpha\beta}$ is
$\epsilon^{0123}=1$. We have introduced upper and lower isospin indices \cite{SZbookchapter,XZThesis}. 
In this section, we focus on the Lagrangian without 
$\Delta$ and study various 
matrix elements: $\bra{N} V^{i}_{\mu},A^{i}_{\mu}, J^{B}_{\mu}
\ket{N}$ and $\bra{N; \pi} V^{i}_{\mu}, A^{i}_{\mu}, J^{B}_{\mu}
\ket{N}$. Definitions of fields and currents 
can be found in Appendix~\ref{app:chisym}. 

\subsection{Power counting and the Lagrangian}
\label{subsec:lagN}

The organization of interaction terms 
is based on power counting \cite{FST97,MCINTIRE07,HU07} and
naive dimensional analysis (NDA) \cite{GEORGI84,GEORGI93}.
We associate with each interaction term an index:
$\hat{\nu} \equiv d+ {n}/{2} + b $.
Here $d$ is the number of derivatives (small momentum transfer) in
the interaction, $n$ is the number of fermion fields, and $b$ is the
number of heavy-meson fields. The Lagrangian is well developed
in Refs.~\cite{tang98,EMQHD07,SZbookchapter,XZThesis}.
We begin with the Lagrangian 
\begin{widetext}
\begin{eqnarray}
 \lag_{N(\hat{\nu} \leqslant 3)}
  &=& \psibar{N}\big(i\ugamma{\mu}[\dcpartial{\mu}
        +ig_{\rho}\rho_{\mu}+ig_{v}V_{\mu}]+g_{A}\ugamma{\mu}\ugammafive\,
        \widetilde{a}_{\mu}  -M+g_{s}\phi\big)N  \notag \\[5pt]
&& {}-\frac{f_{\rho}g_{\rho}}{4M}\, \psibar{N}\rho_{\mu\nu}
        \usigma{\mu\nu}N
        -\frac{f_{v}g_{v}}{4M}\, \psibar{N}V_{\mu\nu} \usigma{\mu\nu}N 
         -\frac{\kappa_{\pi}}{M}\, \psibar{N}\,\widetilde{v}_{\mu\nu}
        \usigma{\mu\nu}N +\frac{4\beta_{\pi}}{M}\, \psibar{N}N \Tr(\widetilde{a}_{\mu}\widetilde{a}^{\mu}) \notag \\[5pt]  
&& {}+\frac{1}{4M}\, \psibar{N} \usigma{\mu\nu}( 2
        \lambda^{(0)}f_{s\mu\nu}+\lambda^{(1)}F^{(+)}_{\mu\nu} ) N 
        +\frac{i\kappa_{1}}{2M^{2}}\, \psibar{N}
        \dgamma{\mu}\overset{\leftrightarrow}{\dcpartial{\nu}} N
        \Tr\left(\widetilde{a}^{\mu}\widetilde{a}^{\nu}\right)   \ . \label{eqn:lagNlowest}
\end{eqnarray}
\end{widetext}
$\dcpartial{\mu}$ is defined in Eq.~(\ref{eqn:allchiral6}),
$\overset{\leftrightarrow}{\dcpartial{\nu}} \equiv \dcpartial{\nu} -
(\overset{\leftarrow}{\partial_{\nu}} - i\widetilde v_{\nu}+
i\vbg_{(s)\nu})$, and the field tensors are $V_{\mu\nu} \equiv
\partial_{\mu} V_{\nu} - \partial_{\nu} V_{\mu}$ and
$\rho_{\mn} \equiv \dcpartial{[\mu}\rho_{\nu
        ]}+i\overline{g}_{\rho}[\rho_{\mu}\,
        , \, \rho_{\nu}] $.
The superscripts ${}^{(0)}$ and ${}^{(1)}$ denote the isospin.
Next is a purely mesonic piece:
\begin{widetext}
\begin{eqnarray}
\lag_{\mathrm{meson} (\hat{\nu} \leqslant4)} &=& \half \,
       \partial_{\mu}\phi\,\partial^{\mu}\phi
       + \frac{1}{4} f^{2}_{\pi} \Tr[\dcpartial{\mu}U(\ucpartial{\mu}U)^{\dagger}] 
       +\frac{1}{4} f^{2}_{\pi}\, m^{2}_{\pi}\Tr(U+U^{\dagger}-2) \notag
       \\[5pt]
&& -\half \Tr(\rho_{\mu\nu}\rho^{\mu\nu}) -\frac{1}{4} \, V^{\mu\nu}V_{\mu\nu}
     +\frac{1}{2g_{\gamma}} \left(
       \Tr(F^{(+)\mu\nu}\rho_{\mu\nu})+\frac{1}{3}\, f_{s}^{\mu\nu}V_{\mu\nu}
       \right)
      \ . \label{eqn:photonmeson}
\end{eqnarray}
\end{widetext}
We only show the kinematic terms and photon
couplings to the vector fields. The latter are used to generate VMD. 
Other $\nu=3$ and $\nu=4$ terms in $\lag_{\mathrm{meson} (\hat{\nu}
\leqslant4)}$ are important for describing the bulk properties of
nuclear many-body systems and can be found in \cite{FST97,FTS95,FST96,SZbookchapter,XZThesis}. 
The only manifest chiral-symmetry breaking is through the nonzero pion mass.
Other chiral-symmetry violating terms and multiple pion interactions are
not considered in this calculation.
Finally, we have
\begin{widetext}
\begin{eqnarray}
\lag_{N,\pi (\hat{\nu} =4)}&=&\frac{1}{2M^{2}}\,
       \psibar{N}\dgamma{\mu}(2\beta^{(0)}
       \partial_{\nu}f_{s}^{\mu\nu}+\beta^{(1)}\dcpartial{\nu}F^{(+)\mu\nu}
       +\beta_{A}^{(1)}\ugammafive \dcpartial{\nu}F^{(-)\mu\nu})N  
-\omega_{1}\Tr(F^{(+)}_{\mu\nu}\, \widetilde{v}^{\mu\nu})  
 +\omega_{2} \Tr(\widetilde{a}_{\mu}\dcpartial{\nu}F^{(-)\mu\nu}) \notag \\[5pt]
  && {}  +\omega_{3} \Tr \left( \widetilde{a}_{\mu} i
       \left[\widetilde{a}_{\nu} \, , \, F^{(+)\mu\nu} \right]\right)      
        -g_{\rho\pi\pi}\frac{2f^{2}_{\pi}}{m^{2}_{\rho}}
       \Tr(\rho_{\mu\nu}\widetilde{v}^{\mu\nu})  
+\frac{c_{1}}{M^{2}}\, \psibar{N}\ugamma{\mu} N
       \Tr \left(\widetilde{a}^{\nu}\, \psibar{F}^{(+)}_{\mu\nu}\right) \notag  \\[5pt]
&& {}  +\frac{e_{1}}{M^{2}}\, \psibar{N}\ugamma{\mu}\,
       \widetilde{a}^{\nu}N \,
       \psibar{f}_{s\mu\nu} 
       +\frac{c_{1\rho}g_{\rho}}{M^{2}}\, \psibar{N}\ugamma{\mu} N\Tr
       \left(\widetilde{a}^{\nu}\, \psibar{\rho}_{\mu\nu}\right)
       +\frac{e_{1v}g_{v}}{M^{2}}\, \psibar{N}\ugamma{\mu}\,
       \widetilde{a}^{\nu}N \, \psibar{V}_{\mu\nu}\ . \label{eqn:c1rhoe1rho}
\end{eqnarray}
\end{widetext}
Note that $\lag_{N,\pi (\hat{\nu} =4)}$ is not a complete list of
all possible $\hat{\nu}=4$ interaction terms. The terms listed in the 
first two rows generate the form factors of currents for nucleons and pions.
$g_{\rho\pi\pi}$ is used for VMD. Special
attention should be given to the $c_{1}, e_{1}, c_{1\rho}$, and
$e_{1\rho}$ couplings, since they are the only relevant
$\hat{\nu}=4$ terms for NC photon production. Further discussion
will be given in Secs.~\ref{subsubsec:diagramef} and
\ref{subsec:ncphotonprod}.

\subsection{ Contributions to current matrix elements from
irreducible diagrams}  \label{sec:currentmatrixelement}

To calculate various current matrix elements, we need to understand
the background fields in terms of electroweak boson fields; this
connection is given in Appendix~\ref{app:chisym}. 
Based on the Lagrangian, we can calculate the matrix elements
$\bra{N} V^{i}_{\mu}, A^{i}_{\mu}$, $J^{B}_{\mu} \ket{N}$ and
$\bra{N; \pi} V^{i}_{\mu}$, $A^{i}_{\mu}$, $J^{B}_{\mu} \ket{N}$
[diagram (f) in Fig.~\ref{fig:feynmanpionproduction}] at
tree level; loops are not
included; only diagrams with contact structure are included\footnote{%
The expressions for the currents listed below differ from those in
Refs.~{\cite{EMQHD07,AXC02}} because contributions from
non-minimal and vector meson-dominance terms are included here.}.
Because of VMD, we can extrapolate the current to nonzero 
$Q^{2}$ \cite{EMQHD07,BDS10}. The results are given
below, and the explicit calculations are shown in
Appendix~\ref{app:ff}. Note that $q^{\mu}$ is defined as the
\emph{incoming} momentum transfer at the vertex; in terms of initial
and final nucleon momenta, $q^{\mu} \equiv p^{\mu}_{nf} -
p^{\mu}_{ni}$. Similarly, $q^{\mu} + p^{\mu}_{ni} = p^{\mu}_{nf} +
k^{\mu}_{\pi}$ for pion production.

First, the matrix elements of the nucleon's vector and baryon current, and 
the axial-vector current in pion production are the following:
\begin{widetext}
\begin{align}
 \bra{N, B} V^{i}_{\mu} \ket{N, A} 
        &= \bra{B} \uhalftau{i}\ket{A}\,
          \psibar{u}_{f}
          \left(\dgamma{\mu}
          +2\delta F_{1}^{V,md}\,\frac{q^{2}\dgamma{\mu}-\slashed{q} q_{\mu}}{q^{2}}
          +2F_{2}^{V,md}\,\frac{\dsigma{\mn}iq^{\nu}}{2M}\right)u_{i} 
\equiv \bra{B} \uhalftau{i}\ket{A}\, \psibar{u}_{f}
          \Gamma_{V\mu}(q) u_{i} \ ,
          \label{eqn:NNvectorcurrentwithff} \\[5pt]
\bra{N, B} J^{B}_{\mu} \ket{N, A}&=  \delta_{B}^{A}\,
          \psibar{u}_{f} \left( \dgamma{\mu}
          +2\delta F_{1}^{S,md}\, \frac{q^{2}\dgamma{\mu}-\slashed{q} q_{\mu}}{q^{2}}
          +2F_{2}^{S,md}\frac{\dsigma{\mn}iq^{\nu}}{2M} \right) u_{i} 
\equiv \delta_{B}^{A}\, \psibar{u}_{f} \Gamma_{B\mu}(q) u_{i} \ ,
          \label{eqn:NNbaryoncurrentwithff}  \\
 \bra{N, B; \pi,j,k_{\pi}} A^{i}_{\mu} & \ket{N, A} =
-\frac{\epsilon^{i}_{\,jk}}{f_{\pi}}\,
          \bra{B}\uhalftau{k}\ket{A}\,
          \psibar{u}_{f}\ugamma{\nu}u_{i} \left[g_{\mn}+2\delta F_{1}^{V,md}((q-k_{\pi})^{2})
          \frac{ q\cdot(q-k_{\pi})g_{\mn}-(q-k_{\pi})_{\mu}q_{\nu}}{(q-k_{\pi})^{2}}
          \right] \notag \\[5pt]
&\hspace{40pt} -\frac{\epsilon^{i}_{\,jk}}{f_{\pi}}\, \bra {B}
\uhalftau{k}\ket{A}\, \psibar{u}_{f}\frac{\dsigma{\mn}iq^{\nu}}{2M}\, u_{i}
          \left[ 2\lambda^{(1)} +2\delta F^{V,md}_{2}((q-k_{\pi})^{2})\,
          \frac{q\cdot(q-k_{\pi})}{(q-k_{\pi})^{2}} \right] 
        \notag\\[5pt]
&  \equiv\frac{\epsilon^{i}_{\,jk}}{f_{\pi}}\, \bra {B}
          \uhalftau{k}\ket{A}\, \psibar{u}_{f} \Gamma_{A\pi \mu}(q,k_{\pi})
          u_{i} \ . \label{eqn:NNpionaxialcurrentwithff}
\end{align}
\end{widetext}
Here $m_{\rho}=0.776\ \mathrm{GeV}$, $m_{v}=0.783 \ \mathrm{GeV}$, 
$\delta F \equiv F(q^{2})-F(0)$ (also true for other form factors), and
\begin{widetext}
\begin{eqnarray}
F_{1}^{V,md}&=&\half \left(1+ \frac{\beta^{(1)}}{M^{2}}\,  q^{2}
          -\frac{g_{\rho}}{g_{\gamma}} \frac{q^{2}}{q^{2}-m^{2}_{\rho}} \right) \ ,  \
          \beta^{(1)}=-1.35, \ \frac{g_{\rho}}{g_{\gamma}}=2.48  \label{eqn:F1vmd}
          \ , \\[5pt]
F_{2}^{V,md}&=&\half
          \left(2\lambda^{(1)}-\frac{f_{\rho}g_{\rho}}{g_{\gamma}}
          \frac{q^{2}}{q^{2}-m^{2}_{\rho}} \right)  \ ,  \
         \lambda^{(1)}=1.85, \, f_{\rho}=3.04 \label{eqn:F2vmd}
          \ , \\[5pt]
F_{1}^{S,md}&=&\half \left(1+ \frac{\beta^{(0)}}{M^{2}}\, q^{2}
          -\frac{2 g_{v}}{3 g_{\gamma}} \frac{q^{2}}{q^{2}-m^{2}_{v}} \right) \ , \
         \beta^{(0)}=-1.40, \ \frac{g_{v}}{ g_{\gamma}}=3.95 \label{eqn:F1smd}
          \ , \\[5pt]
F_{2}^{S,md}&=&\half \left(2\lambda^{(0)}-\frac{2
          f_{v}g_{v}}{3g_{\gamma}}
          \frac{q^{2}}{q^{2}-m^{2}_{v}} \right) \ , \
          \lambda^{(0)}=-0.06, \ f_{v}=-0.19 \label{eqn:F2smd}
          \ .
\end{eqnarray}          
\end{widetext}

We can also use this procedure to expand the axial-vector current 
in powers of $q^2$ using the
Lagrangian constants $g_{A}$ and $\beta_{A}^{(1)}$. In fact, we can
improve on this by including the axial-vector meson ($a_{1\mu}$)
contribution to the matrix elements, which would arise from the
interactions: $g_{a_{1}}\psibar{N}\ugamma{\mu}\ugammafive a_{1\mu}
N$ and $c_{a_{1}}\Tr \left(F^{(-)\mn} a_{1\mn} \right)$. Here
$a_{1\mu}=a_{1i\mu}\tau^{i}/2$ and $a_{1\mn}\equiv\dcpartial{\mu}
a_{1\nu}-\dcpartial{\nu} a_{1\mu}$, where $a_{1i\mu}$ are the fields
of the $a_{1}$ meson (with its mass denoted as $m_{a_{1}}=1.26 \
\mathrm{GeV}$). Then we obtain
\begin{widetext}
\begin{eqnarray}
\bra{N, B} A^{i}_{\mu} \ket{N, A} &=& -G_{A}^{md}(q^{2})\,
          \bra{B} \uhalftau{i}\ket{A}\, \psibar{u}_{f} \left(
          \dgamma{\mu}-\frac{q_{\mu}\slashed{q}}{q^{2}-m_{\pi}^{2}}\right)
          \ugammafive u_{i} 
\equiv \bra{B} \uhalftau{i}\ket{A}\, \psibar{u}_{f}\Gamma_{A \mu}
(q) u_{i}\ ,
          \label{eqn:NNaxialcurrentff}  \\[5pt]%
\bra{N, B,\pi, j}  V^{i}_{\mu} \ket{N, A}&=&
\frac{\epsilon^{i}_{\,jk}}{f_{\pi}}\,
          \bra {B} \uhalftau{k}\ket{A}\, \psibar{u}_{f}\bigg( G_{A}^{md}(0)
          \dgamma{\mu}\ugammafive 
          +\delta G_{A}^{md}((q-k_{\pi})^{2})\,
          \frac{q\cdot (q-k_{\pi})g_{\mn}-(q-k_{\pi})_{\mu}
          q_{\nu}}{(q-k_{\pi})^{2}}\,
          \ugamma{\nu}\ugammafive \bigg) u_{i}   \notag \\[5pt]
&\equiv& \frac{\epsilon^{i}_{\,jk}}{f_{\pi}}\,
          \bra {B} \uhalftau{k}\ket{A}\, \psibar{u}_{f} \Gamma_{V\pi \mu}(q,k_{\pi})u_{i}
          \ , \label{eqn:NNpionvectorcurrentff} \\
G_{A}^{md}(q^{2})&\equiv&
          g_{A}-\beta_{A}^{(1)}\,\frac{q^{2}}{M^{2}}
          - \frac{2c_{a_{1}}g_{a_{1}}q^{2}}{q^{2}-m_{a_{1}}^{2}}\ ,  \
g_{A}=1.26, \ \beta_{A}^{(1)}=2.27, \ c_{a_{1}}g_{a_{1}}=3.85\ .
          \label{eqn:defofGA1} 
\end{eqnarray}
\end{widetext}

For the pion's vector current form factor  \cite{FST97},
\begin{widetext}
\begin{eqnarray}
\bra{\pi,k,k_{\pi}} V^{i}_{\mu}\ket{\pi,j,k_{\pi}-q} &=&
i\epsilon^{ij}_{\;\; k}\,
          \left[(2k_{\pi}-q)_{\mu} +2 \delta F_{\pi}^{md}(q^{2}) \left(k_{\pi \mu}
          -\frac{q\cdot k_{\pi}}{q^{2}}\, q_{\mu}\right)\right] 
           \equiv  i\epsilon^{ij}_{\;\; k}\, P_{V \mu}(q,k_{\pi}) \ ,\notag  \\[5pt]
 \qquad   F_{\pi}^{md}(q^{2})
         &\equiv&
        \left(1-\frac{g_{\rho\pi\pi}}{g_{\gamma}}\frac{q^{2}}{q^{2}-m^{2}_{\rho}}\right)
         \ , \ 
          \frac{g_{\rho\pi\pi}}{g_{\gamma}}=1.20 \ . \label{eqn:pionvectorcurrentff}  
\end{eqnarray}
\end{widetext}

To determine the couplings in Eqs.~(\ref{eqn:F1vmd}),
(\ref{eqn:F2vmd}), (\ref{eqn:F1smd}), (\ref{eqn:F2smd}),
(\ref{eqn:defofGA1}) and (\ref{eqn:pionvectorcurrentff}), we compare our results
with the fitted form factors
\cite{FST97,kelly04}. We require that the behavior of our vector- and 
baryon-meson-dominance form factors near
$Q^{2}=0$ be close to that of the fitted form factors \cite{kelly04}. The
nucleon's axial-vector current used to fit our
$G_{A}^{md}$ is parametrized as
$G_{A}(q^{2})=g_{A}/(1-q^{2}/M_{A}^{2})^{2}$ with $g_{A}=1.26$ and
$M_{A}=1.05 \, \mathrm{GeV}$ \cite{EW88}. As shown in Ref.~\cite{BDS10}, the
form factors due to vector meson dominance become inadequate at
$Q^{2} \approx 0.3 \, \mathrm{GeV}^{2}$. This is also true for the
axial-vector's parametrization. This indicates that the EFT Lagrangian is
only applicable for $E_{l} \leqslant 0.5 \, \mathrm{GeV}$ in
lepton--nucleon interactions, above which $Q^{2}$ exceeds the limit.
This will be clarified in the kinematical analysis of
Sec.~\ref{subsec:kinematics}.

\section{Lagrangian involving $\Delta (1232)$} \label{sec:LDelta}

\subsection{Lagrangian} \label{subsubsec:deltaprop}
Two remarks are in order here \cite{SZbookchapter,XZThesis}:
First, the theory is self-consistent with general interactions involving $\psi^{\mu}$;
second, the so-called off-shell couplings, which have the
form $\gamma_{\mu} \psi^{\mu}$, $\dcpartial{\mu} \psi^{\mu},
\overline{\psi}^{\mkern3mu\mu} \gamma_{\mu},$ and $\dcpartial{\mu}
\overline{\psi}^{\mkern3mu\mu}$, can be considered as redundant.
For the chiral symmetry realization, 
$\Delta^{\ast a}$ belong to an $I=3/2$ multiplet [$a= (\pm 3/2, \pm 1/2)$]. 
Moreover in the power counting 
of vertices, the $\Delta$ is counted in the same way as nucleons.

Consider first $\lag_{\Delta}\ (\hat{\nu}\leqslant
3)$:
\begin{widetext}
\begin{eqnarray}
\lag_{\Delta}&=&\frac{-i}{2}\,
          \psibar{\Delta}^{\mkern4mu a}_{\mu}\{\usigma{\mn}\, , \, (i\slashed{\ucpartial{}}
          -h_{\rho}\slashed{\rho}-h_{v}\slashed{V}-m
          +h_{s}\phi)\}_{a}^{\mkern3mu b}\, \Delta_{b\nu}
          + \widetilde{h}_{A}\psibar{\Delta}^{\mkern4mu a}_{\mu}
          \slashed{\widetilde{a}}_{a}^{\mkern3mu b} \ugammafive \Delta^{\mu}_{b}
          \notag \\[5pt]
&&{}-\frac{\widetilde{f}_{\rho}h_{\rho}}{4m}\,
          \psibar{\Delta}_{\lambda}\,
          \rho_{\mn}\usigma{\mn}\Delta^{\lambda}
          -\frac{\widetilde{f}_{v}h_{v}}{4m}\,
          \psibar{\Delta}_{\lambda} V_{\mn}\usigma{\mn}\Delta^{\lambda}           
          -\frac{\widetilde{\kappa}_{\pi}}{m}\,
          \psibar{\Delta}_{\lambda}\widetilde{v}_{\mn}\usigma{\mn}\Delta^{\lambda}
          +\frac{4\widetilde{\beta}_{\pi}}{m}\,
          \psibar{\Delta}_{\lambda}\Delta^{\lambda}\Tr(\widetilde{a}^{\mu}\,
          \widetilde{a}_{\mu}) \ . \label{eqn:deltalaglowest}  
\end{eqnarray}     
\end{widetext}
This is essentially a copy of the corresponding Lagrangian for nucleons.

To produce the $N \leftrightarrow \Delta$ transition currents, we
construct the following Lagrangians ($\hat{\nu}\leqslant4$):

\begin{widetext}
\begin{eqnarray}
\lag_{\Delta,N,\pi}&=&h_{A}\psibar{\Delta}^{\mkern4mu a
          \mu}\, \Tdagger{iA}{a}\, \widetilde{a}_{i\mu}N_{A} +C.C. \ ,  \\[5pt]
\lag_{\Delta,N,\mathrm{bg}}&=& \frac{ic_{1\Delta}}{M}\,
          \psibar{\Delta}^{\mkern4mu a}_{\mu}\dgamma{\nu}\ugammafive\,
          \Tdagger{iA}{a}F_{i}^{(+)\mn}N_{A}
          +\frac{ic_{3\Delta}}{M^{2}}\,
          \psibar{\Delta}^{\mkern4mu a}_{\mu}\, i\ugammafive\, \Tdagger{iA}{a}
          (\dcpartial{\nu}F^{(+)\mn})_{i} N_{A}  
          +\frac{c_{6\Delta}}{M^{2}}\, \psibar{\Delta}^{\mkern4mu
          a}_{\lambda}
          \dsigma{\mu\nu} \Tdagger{iA}{a}
          (\ucpartial{\lambda}\psibar{F}^{(+)\mn})_{i} N_{A} \notag \\[5pt]
&&{}-\frac{d_{2\Delta}}{M^{2}}\, \psibar{\Delta}^{\mkern4mu
          a}_{\mu}\,
          \Tdagger{iA}{a} (\dcpartial{\nu}F^{(-)\mn})_{i}N_{A}
          -\frac{id_{4\Delta}}{M}\, \psibar{\Delta}^{\mkern4mu a}_{\mu} \dgamma{\nu}\,
          \Tdagger{iA}{a} F_{i}^{(-)\mn}N_{A} 
          -\frac{id_{7\Delta}}{M^{2}}\, \psibar{\Delta}^{\mkern4mu
          a}_{\lambda}
          \dsigma{\mn}\Tdagger{iA}{a}(\ucpartial{\lambda}F^{(-)\mn})_{i}N_{A} \notag \\
&&{}+ C.C. \ , \label{eqn:truelagdeltaNbackground} \\[5pt]
\lag_{\Delta,N,\rho}&=&\frac{ic_{1\Delta\rho}}{M}\,
          \psibar{\Delta}^{\mkern4mu a}_{\mu}\,\dgamma{\nu}\ugammafive\,
          \Tdagger{iA}{a}\rho_{i}^{\mn}N_{A}
          +\frac{ic_{3\Delta\rho}}{M^{2}}\, \psibar{\Delta}^{\mkern4mu a}_{\mu}\,
          i\ugammafive\, \Tdagger{iA}{a} (\dcpartial{\nu}\rho^{\mn})_{i} N_{A}
          +\frac{c_{6\Delta\rho}}{M^{2}}\, \psibar{\Delta}^{\mkern4mu
          a}_{\lambda} \dsigma{\mu\nu}\,
          \Tdagger{iA}{a} (\ucpartial{\lambda}\, \psibar{\rho}^{\mkern2mu\mn})_{i} N_{A} \notag \\
&&{}+ C.C. \ . \label{eqn:truelagdeltaNrho}
\end{eqnarray}
\end{widetext}
Here $T^{\dagger
\,\,iA}_{a} = \langle \frac{3}{2}; a \vert 1,\frac{1}{2};i,A \rangle$, which are (complex conjugate of) Clebsch-Gordan coefficients. 

\subsection{Transition currents} \label{subsubsec:tc}

We can express the transition current's matrix element as follows:
\begin{align}
\bra{\Delta, a,p_{\Delta}} V^{i\mu} & (A^{i\mu}) \ket{N, A,p_{N}} \equiv \notag \\
 {} & \Tdagger{iA}{a}\, \psibar{u}_{\Delta\alpha}(p_{\Delta})\,
          \Gamma_{V(A)}^{\alpha\mu}(q)\, u_{N}(p_{N})\ . 
          \label{eqn:transitioncurrentvertex} 
\end{align}
Based on the Lagrangians, we find (noting that
$\sigma_{\mn} \epsilon^{\mu\nu\alpha\beta} \propto i
\sigma^{\alpha\beta} \ugammafive$)
\begin{widetext}
\begin{eqnarray}
\Gamma_{V}^{\alpha\mu}&=&\frac{2c_{1\Delta}(q^{2})}{M}\,
          (q^{\alpha}\ugamma{\mu}-\slashed{q}g^{\alpha\mu}) \ugammafive
          +\frac{2c_{3\Delta}(q^{2})}{M^{2}}\, (q^{\alpha}q^{\mu}-g^{\alpha\mu}q^{2})
          \ugammafive 
          -\frac{8c_{6\Delta}(q^{2})}{M^{2}}\,
          q^{\alpha}\usigma{\mn}iq_{\nu}\ugammafive \ , \\ [5pt]
\Gamma_{A}^{\alpha\mu}&=&-h_{A} \left(
          g^{\alpha\mu}-\frac{q^{\alpha}q^{\mu}}{q^{2}-m^{2}_{\pi}} \right)
          +\frac{2d_{2\Delta}}{M^{2}}\, (q^{\alpha}q^{\mu}-g^{\alpha\mu}q^{2}) 
          -\frac{2d_{4\Delta}}{M}\, (q^{\alpha}\ugamma{\mu}-g^{\alpha\mu}\slashed{q})          
-\frac{4d_{7\Delta}}{M^{2}}\, q^{\alpha}\usigma{\mn}iq_{\nu}\ , \\[5pt]
c_{i\Delta}(q^{2}) &\equiv& c_{i\Delta}
          +\frac{c_{i\Delta\rho}}{2g_{\gamma}}\, \frac{q^{2}}{q^{2}-m^{2}_{\rho}}\ ,
         \, \, i=1, 3, 6, \notag \\[5pt]
c_{1\Delta}&=&1.21 \ , \quad c_{3\Delta}=-0.61 \ , \quad c_{6\Delta}=-0.078 \ , \  
\frac{c_{1\Delta\rho}}{g_{\gamma}}=-4.58 \ , \quad
\frac{c_{3\Delta\rho}}{g_{\gamma}}=2.32 \ , \quad
\frac{c_{6\Delta\rho}}{g_{\gamma}}=0.30 \ . \label{eqn:cimd2}
\end{eqnarray}
\end{widetext}
Similar to the $c_{i \Delta}(q^{2})$, we can introduce
axial-vector meson exchange into the axial transition current, which
leads to a structure for the $d_{i \Delta}(q^{2})$ similar
to that of the $c_{i \Delta}(q^{2})$. There is one subtlety
associated with the realization of $h_{A}(q^{2})$: 
with our Lagrangian, we have a pion-pole
contribution associated with the $h_{A}$ coupling, and all the
higher-order terms contained in $\delta h_{A}(q^{2})\equiv
h_{A}(q^{2})-h_{A}$ conserve the axial transition current. With the
limited information about manifest chiral-symmetry breaking, we 
ignore this subtlety and still use the form of the
$c_{1\Delta}(q^{2})$ to parametrize $h_{A}(q^{2})$:
\begin{widetext}
\begin{align}
h_{A}(q^{2}) &\equiv h_{A}+h_{\Delta a_{1}} \,  \frac{q^{2}}{q^{2}-m^{2}_{a_{1}}}\ ,  \quad
h_{A}=1.40 \ , \quad h_{\Delta a_{1}}=-3.98 \ ,  \label{eqn:dimd0} \\[5pt]      
d_{i\Delta}(q^{2}) &\equiv d_{i\Delta}
          + d_{i \Delta a_{1}}\, \frac{q^{2}}{q^{2}-m^{2}_{a_{1}}}\ ,
          \, \, i=2, 4, 7, \notag \\[5pt]
d_{2\Delta}&=-0.087, \quad d_{4\Delta}=0.20,
          \quad d_{7\Delta}=-0.04 \ ,   \quad
d_{2 \Delta a_{1}}=0.25 \ , \quad
          d_{4 \Delta a_{1}}=-0.58 \ , \quad d_{7 \Delta a_{1}}=0.12 \ .
          \label{eqn:dimd2} 
\end{align}
\end{widetext}

To determine the coefficients in the transition form factors shown
in Eqs.~(\ref{eqn:cimd2}) (\ref{eqn:dimd0})  
and (\ref{eqn:dimd2}), we compare ours
with one of the conventional form factors used in the literature. In
Refs.~\cite{HERNANDEZ07, GRACZYK09} for example, the definition for 
$\bra{\Delta,\half} j_{cc+}^{\mu} \ket{N,-\half}$ [$=-\sqrt{{2}/{3}}\
          \psibar{u}_{\alpha}(p_{\Delta})\left(\Gamma_{V}^{\alpha\mu}
          +\Gamma_{A}^{\alpha\mu}\right)
          u(p_{N})$] is
\begin{widetext}
\begin{eqnarray}
&& \psibar{u}_{\alpha}(p_{\Delta}) \left\{ \left[
          \frac{C_{3}^{V}}{M}\, (g^{\alpha\mu} \slashed{q}-q^{\alpha}
          \ugamma{\mu}) +\frac{C_{4}^{V}}{M^{2}}\, (q\cdot
          p_{\Delta}\, g^{\alpha\mu}-q^{\alpha} p_{\Delta}^{\mu}) + 
           \frac{C_{5}^{V}}{M^{2}}\, (q\cdot p_{N}\,
          g^{\alpha\mu}-q^{\alpha}
          p_{N}^{\mu}) \right]  \right. \ugammafive  \notag \\[5pt]
&&+ \left. \left[ \frac{C_{3}^{A}}{M}\,
          (g^{\alpha\mu}\slashed{q}-q^{\alpha}\ugamma{\mu})
          +\frac{C_{4}^{A}}{M^{2}}\, (q\cdot p_{\Delta}\, g^{\alpha\mu}-q^{\alpha}
          p_{\Delta}^{\mu}) + C_{5}^{A}g^{\alpha\mu}+\frac{C_{6}^{A}}{M^{2}}\,
          q^{\mu}q^{\alpha} \right] \right\} u(p_{N})  \ .  
\end{eqnarray}
\end{widetext}
We use the ``Adler parametrization'' \cite{ADLER68} in
Ref.~\cite{GRACZYK09} to fit our meson-dominance form factors.
Now supposing the baryons are on shell, we can represent 
the conventional basis as linear combinations of our basis. 
For example,
\begin{align}
& q^{\alpha}\usigma{\mn}iq_{\nu}  \ugammafive 
= (m-M) (q^{\alpha} \ugamma{\mu}-g^{\alpha\mu}\slashed{q})\ugammafive  \notag \\[5pt]
&-(q^{\alpha} p_{\Delta}^{\mu}-q\cdot p_{\Delta}g^{\alpha\mu})\ugammafive 
- (q^{\alpha} p_{N}^{\mu}-q\cdot p_{N}g^{\alpha\mu})\ugammafive  \ .  \label{eqn:gammaVonshell} 
\end{align}
A similar relation holds 
with $\ugammafive$ deleted on both sides and $(m-M)$ changed to $(m+M)$. 
We can obtain the relation between form factors associated with the 
two bases:
\begin{align}
c_{1\Delta}&=\sqrt{\frac{3}{2}}\,\left[\frac{C_{3}^{V}}{2}+\frac{m-M}{2M}\frac{(C_{4}^{V}
          +C_{5}^{V})}{2}\right] \ ,
\label{eqn:axialtransitionff1}
          \\[5pt]
c_{3\Delta}&=\sqrt{\frac{3}{2}}\ \frac{(C_{4}^{V}-C_{5}^{V})}{4} \ , \quad
c_{6\Delta}=\sqrt{\frac{3}{2}}\ \frac{(C_{4}^{V}+C_{5}^{V})}{16} \ , 
\label{eqn:axialtransitionff2}
          \\[5pt]
h_{A}&=\sqrt{\frac{3}{2}}\ C_{5}^{A} \ , \hspace{47pt}
d_{2\Delta}=\sqrt{\frac{3}{2}}\ \frac{C_{4}^{A}}{4}  , 
\label{eqn:axialtransitionff3}
          \\[5pt]
d_{4\Delta}&=-\sqrt{\frac{3}{2}} \left(\frac{C_{3}^{A}}{2}
          +\frac{m+M}{2M} \frac{C_{4}^{A}}{2}\right) \ , 
d_{7\Delta}=\sqrt{\frac{3}{2}}\ \frac{C_{4}^{A}}{8} \ . 
\label{eqn:axialtransitionff}
\end{align}
We assume that these relations hold away from the resonance. 
It can be shown that, at low energy, the
differences in observables due to using the two bases, with these
relations applied, are negligible. 
Moreover, the $q^{2}$ dependence of these $c_{i\Delta}$
and $d_{i\Delta}$ form factors can be realized in terms of meson
dominance. We then require that the meson-dominance form factors be
as close as possible to the ones indicated in 
Eqs.~(\ref{eqn:axialtransitionff1}) to (\ref{eqn:axialtransitionff}), 
and we get the couplings shown in 
Eqs.~(\ref{eqn:cimd2}), (\ref{eqn:dimd0}),
and (\ref{eqn:dimd2}). 
However, we should expect the leading-order meson-dominance expressions 
would fail above $Q^{2} \approx 0.3 \, \mathrm{GeV}^{2}$.  

\section{Feynman diagrams}
\label{sec:diag}

\begin{figure}
\centering
\includegraphics[scale=0.5]{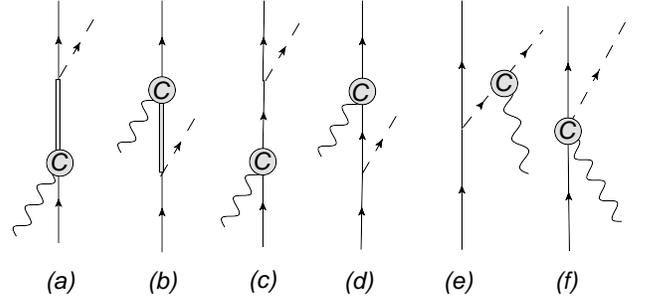}
\caption{Feynman diagrams for
pion production. Here, {\bf C} stands for various types of currents
including vector, axial-vector, and baryon currents. Some diagrams
may be zero for some specific type of current. For example, diagrams
(a) and (b) will not contribute for the (isoscalar) baryon current.
Diagram (e) will be zero for the axial-vector current. The pion-pole
contributions to the axial current in diagrams (a) (b) (c) (d)
and (f) are included in the vertex functions of the currents.}
\label{fig:feynmanpionproduction}
\end{figure}

Tree-level Feynman diagrams for pion production due to the vector
current, axial-vector current, and baryon current are shown in
Fig.~\ref{fig:feynmanpionproduction}. In this section, we 
calculate different matrix elements for pion production and
photon production. The Feynman diagrams for photon production can be
viewed as diagrams in Fig.~\ref{fig:feynmanpionproduction} with an
outgoing $\pi$ line changed to a $\gamma$ line. It turns out
that diagram (e) in Fig.~\ref{fig:feynmanpionproduction} is negligible in
NC photon production, since it is associated with $1-4 \sin^{2}\theta_{w}$ \cite{RHill09}.

First let us outline the calculation of the interaction amplitude
$M$. Consider CC pion production (in the one-weak-boson-exchange
approximation):
\begin{eqnarray}
M&=&4\sqrt{2}\, G_{F} V_{ud}\, \langle J_{L i \mu}^{(lep)} \rangle
\langle J_{L}^{(had) i \mu} \rangle_{\pi}\ .
\label{eqn:amplitudedefccpion}
\end{eqnarray}
where $i=+1,-1$. In Eq.~(\ref{eqn:amplitudedefccpion}), $G_{F}$ is
the Fermi constant, $V_{ud}$ is the CKM matrix element corresponding
to $u$ and $d$ quark mixing,
and $ \langle J_{L}^{(had) i \mu} \rangle_{\pi} \equiv \bra{N B, \pi
j } J_{L}^{i \mu }\ket{N A}$ (The definitions of currents can be
found in Appendix~\ref{app:chisym}.)
$ \langle J_{L i \mu}^{(lep)}\rangle
\equiv \bra{l(\bar{l})} J_{L i \mu} \ket{\nu_{l} (\bar{\nu_{l}})}$
is the well-known leptonic-charged-current matrix element.
For NC pion production, we need to set $V_{ud}=1$,
$\langle J_{L}^{(had) i \mu} \rangle_{\pi}\rightarrow \langle
J_{NC}^{(had)  \mu} \rangle_{\pi}$,
and $\langle J_{L i \mu}^{(lep)} \rangle \rightarrow \langle J_{NC  \mu}^{(lep)} \rangle$
in Eq.~(\ref{eqn:amplitudedefccpion}).
Here $ \langle J_{NC \mu}^{(lep)} \rangle$ is the 
leptonic-neutral-current matrix element, and  $\langle J_{NC}^{(had)
\mu} \rangle_{\pi} \equiv \bra{N B, \pi j } J_{NC}^{ \mu }\ket{N
A}$. For NC photon production, we have an expression similar to that of 
NC pion production with 
$ \langle J_{NC}^{(had) \mu} \rangle_{\pi} \rightarrow \langle
J_{NC}^{(had)  \mu} \rangle_{\gamma}$,
while $ \langle J_{NC}^{(had)  \mu} \rangle_{\gamma} \equiv \bra{N
B, \gamma } J_{NC}^{ \mu }\ket{N A}$.

Now consider the power counting for $ \langle J^{(had) \mu}
\rangle_{\pi(\gamma)}$ in
Eq.~(\ref{eqn:amplitudedefccpion}) for various processes. 
The order of the diagram
$(\nu)$ is counted as \cite{tang98} $\nu=2L+2-
 E_{n}/2 +\sum_{i} \#_{i}(\hat{\nu}_{i}-2)$, where $L$
is the number of loops, $E_{n}$ is the number of external baryon
lines, $\hat{\nu}_{i}\equiv d_{i}+n_{i}/{2}+b_{i}$ is the
order of the vertex $(\hat{\nu})$ mentioned in
Sec.~\ref{subsec:lagN}, and $\#_{i}$ is the number of times that
particular vertex appears.
 However, there is a subtlety related with power counting of diagrams
with $\Delta$, which has been carefully discussed in Ref.~\cite{Pascalutsa08}.
Compared to the normal power counting mentioned above, in which the 
baryon propagator scales as ${1}/{O(Q)}$, for diagrams
involving one $\Delta$ in the $s$ channel, we take $\nu \to \nu -1$
in the resonance regime and $\nu \to \nu+1 $ away from the
resonance. Details can be found in Appendix~\ref{app:deltapowercounting}.

Finally, conservation of vector current, conservation of baryon current, and partial conservation of axial-vector current can be easily checked 
for the matrix elements shown in the following. 

\subsection{Diagram (a) and (b)}
Diagram (a) and (b) in Fig.~\ref{fig:feynmanpionproduction} lead to 
currents:
\begin{widetext}
\begin{eqnarray}
\langle V^{i \mu} (A^{i \mu}) \rangle_{\pi}= -\frac{ih_{A}}{f_{\pi}}\,
\T{a}{Bj}\, \Tdagger{iA}{a}\, \psibar{u}_{f} k_{\pi}^{\lambda}\,
S_{F\lambda \alpha}(p)\, \Gamma_{V (A)}^{ \alpha \mu}(p;q,p_{i}) u_{i} -\frac{ih_{A}}{f_{\pi}}\,
          \T{ai}{B}\, \Tdagger{A}{ja}\, \psibar{u}_{f} \psibar{\Gamma}_{V (A)}^{\mu
          \alpha }(p_{f};q,p) S_{F\alpha \lambda }(p)\,
          k_{\pi}^{\lambda}\, u_{i} \ , 
 \label{eqn:vcdelta}
\end{eqnarray}
\end{widetext}
where $k_{\pi}$ is the \emph{outgoing} pion's momentum. 
$\Gamma_{V(A)}^{ \alpha \mu}(p;q,p_{i})$ are defined in
Eq.~(\ref{eqn:transitioncurrentvertex}) while  
$\Delta$'s momentum is $p=q+p_{i}$. 
$\psibar{\Gamma}_{V(A)}^{ \mu \alpha}(p_{f};q,p) \equiv \ugamma{0}
\Gamma_{V(A)}^{\dagger \alpha \mu}(p;-q,p_{f})\ugamma{0}$ while
$p=-q+p_{f}$. In the following, we always have this definition 
of $\psibar{\Gamma}$.
The $\Delta$'s propagator, $S_{F\mn }(p)$, is shown in 
Appendix~\ref{app:deltaprop}.
The subscript $j$ denotes the
isospin of the outgoing pion. For vector current, 
in diagram (a) $\nu_{nr} \geqslant 3$ in the lower-energy region
and $\nu_{r} \geqslant 1$ in the resonance region; in diagram (b)
$\nu_{nr} \geqslant 3$. 
For axial-vector current, in diagram (a) 
$\nu_{nr} \geqslant 2, \nu_{r} \geqslant 0$; in diagram (b) 
$\nu_{nr}  \geqslant 2$. In the power counting, 
the higher-order terms in $\nu$ come from including form factors at the
vertices.
Moreover, the baryon current matrix element is zero $( \langle
J_{B}^{\mu}\rangle_{\pi}=0 )$ in both diagrams.

Now we examine the NC matrix element $\langle J_{NC}^{(had)  \mu}
\rangle_{\gamma}$. First, based on the relations given in
Eq.~(\ref{eqn:ncdef}), we define
\begin{align}
\Gamma_{N}^{\alpha\mu}(p;q,p_{i}) \equiv \left(\half-\sin^{2}\theta_{w}\right)
          \Gamma_{V}^{\alpha\mu}(p;q,p_{i})+\half\, \Gamma_{A}^{\alpha\mu}(p;q,p_{i})\ ,  
\end{align}
Then we find
\begin{widetext}
\begin{eqnarray}
\langle J_{NC}^{\mu}\rangle _{\gamma}&=& e \T{a}{0B}\,
          \Tdagger{0A}{a}\, \psibar{u}_{f}\, \epsilon_{\lambda}^{\ast}(k)
          \psibar{\Gamma}_{V}^{\lambda \alpha}(p_{f};-k,p) S_{F\alpha\beta}(p)
          \Gamma_{N}^{\beta \mu}(p;q,p_{i}) u_{i}  \notag \\[5pt]
&&+e \T{a0}{B}\,
          \Tdagger{A}{a0}\, \psibar{u}_{f} \psibar{\Gamma}_{N}^{\mu \alpha
          }(p_{f};q,p) S_{F\alpha\beta}(p)  \Gamma_{V}^{\beta \lambda
          }(p;-k,p_{i})\, \epsilon_{\lambda}^{\ast}(k)\,  u_{i}   \ ,       
          \label{eqn:ncdelta}
\end{eqnarray}
\end{widetext}
where $k$ is the outgoing photon's momentum and
$\epsilon_{\lambda}^{\ast}(k)$ is its polarization.
For the vector current in the NC, in diagram (a) 
$\nu_{nr} \geqslant 4$, $\nu_{r} \geqslant 2$; in diagram (b)
$\nu_{nr} \geqslant 4$.
For the axial-vector current, in diagram (a) 
$\nu_{nr} \geqslant 3$, $\nu_{r} \geqslant
1$; in diagram (b) $\nu_{nr} \geqslant 3$.

\subsection{Diagrams (c) and (d)}

These two diagrams lead to currents:
\begin{widetext}
\begin{eqnarray}
\langle V^{i \mu} (A^{i \mu})\rangle_{\pi} =-\frac{ig_{A}}{f_{\pi}}\,
          \bra{B} \dhalftau{j}\uhalftau{i}\ket{A}\, \psibar{u}_{f}
          \slashed{k_{\pi}}\ugammafive S_{F}(p) \Gamma_{V(A)}^{\mu}(q) u_{i} -\frac{ig_{A}}{f_{\pi}}\, \bra{B}
          \uhalftau{i}\dhalftau{j}\ket{A}\, \psibar{u}_{f}\Gamma_{V(A)}^{\mu}(q)
          S_{F}(p) \slashed{k_{\pi}}\ugammafive u_{i}\ .
          \label{eqn:vcn}
\end{eqnarray}
\end{widetext}
For the nucleon propagator, 
$p=q+p_{i}$ in diagram (c) and 
$p=-q+p_{f}$ in diagram (d).
$\Gamma_{V(A)}^{\mu}(q)$ has been defined in
Eq.~(\ref{eqn:NNvectorcurrentwithff}). For both currents in both 
diagrams $\nu \geqslant 1$. 
For the baryon current we just need to change 
$\uhalftau{i} \Gamma_{V}^{\mu}(q)$ to 
$\Gamma_{B}^{\mu}(q)$ in Eq.~(\ref{eqn:vcn}), and 
 $\nu\geqslant 1$.

For NC photon production, we get
\begin{widetext}
\begin{eqnarray}
\langle J_{NC}^{\mu}\rangle _{\gamma}&=& e\, \psibar{u}_{f}\,
          \epsilon^{\ast}_{\lambda}(k) \left( \left(\uhalftau{0}\right)_{B}^{\;C}\,
          \Gamma_{V }^{\lambda}(-k)
          + \frac{\delta_{B}^{\;C}}{2}\, \Gamma_{B }^{\lambda}(-k) \right) S_{F}(p)  \notag
          \\[5pt]
&&{}\times \left(\left(\uhalftau{0}\right)_{C}^{\; A}\left[\left(\half - \sin^{2}{\theta_{w}}\right)
          \Gamma_{V}^{\mu}(q)+\half\Gamma_{A}^{\mu}(q)\right]
          - \frac{\delta_{C}^{\; A}}{2}\sin^{2}{\theta_{w}}\, \Gamma_{B}^{\mu}(q)
          \right) u_{i} \notag
          \\[5pt]
&&{}+ e \, \psibar{u}_{f} \left(\left(\uhalftau{0}\right)_{B}^{\; C}\left[\left(\half - \sin^{2}{\theta_{w}}\right)
          \Gamma_{V}^{\mu}(q)+\half\Gamma_{A}^{\mu}(q)\right]
          - \frac{\delta_{B}^{\; C}}{2}\sin^{2}{\theta_{w}} \Gamma_{B}^{\mu}(q) \right) \notag
          \\[5pt]
&&{}\times S_{F}(p)\,  \epsilon^{\ast}_{\lambda}(k) \left(
          \left(\uhalftau{0}\right)_{C}^{\; A}\, \Gamma_{V }^{\lambda}(-k) +
          \frac{\delta_{C}^{\; A}}{2}\, \Gamma_{B }^{\lambda}(-k) \right)
          u_{i}\ ,
\label{eqn:ncn}
\end{eqnarray}
\end{widetext}
where we use the shorthand
$(\uhalftau{0})_{B}^{\; A} = \bra{B} \uhalftau{0}\ket{A}$.
For all three currents, power counting gives $\nu \geqslant
1$. However, this naive power counting does not give an accurate
comparison between the $\Delta$ contributions and the $N$
contributions at low energies, as we discuss later.

\subsection{Diagrams (e) and (f)}
\label{subsubsec:diagramef}

The two diagrams lead to a vector current
\begin{widetext}
\begin{eqnarray}
\langle V^{i \mu}\rangle_{\pi} &=& \frac{g_{A}}{f_{\pi}}\,
          \epsilon^{i}_{\;jk} \bra {B}\uhalftau{k}\ket{A} \,
          \frac{P_{V}^{\mu}(q,k_{\pi})}{(q-k_{\pi})^{2}-m_{\pi}^{2}}\,
          \psibar{u}_{f} (\slashed{q}-\slashed{k_{\pi}}) \ugammafive \, u_{i}  +\frac{\epsilon^{i}_{\,jk}}{f_{\pi}}\, \bra {B}
          \uhalftau{k}\ket{A}\, \psibar{u}_{f} \Gamma_{V\pi
          }^{\mu}(q,k_{\pi})\, u_{i} \ .
          \label{eqn:vcpfcontact}
\end{eqnarray}
\end{widetext}
Here, $P_{V}^{\mu}(q,k_{\pi})$ is defined in
Eq.~(\ref{eqn:pionvectorcurrentff}), $\Gamma_{V\pi
}^{\mu}(q,k_{\pi})$ is defined in
Eq.~(\ref{eqn:NNpionvectorcurrentff}), and $\nu \geqslant 1 $.

For the axial-vector current, diagram (e) does not contribute, and
we find
\begin{widetext}
\begin{eqnarray}
\langle A^{i \mu}\rangle_{\pi}
          &=&\frac{\epsilon^{i}_{\,jk}}{f_{\pi}}\, \bra {B}
          \uhalftau{k}\ket{A}\, \psibar{u}_{f} \Gamma_{A\pi}^{\mu}(q,k_{\pi})\, u_{i}
          +\frac{\epsilon^{i}_{\;jk}}{f_{\pi}}\,
          \bra{B}\uhalftau{k}\ket{A}\,
          \frac{q^{\mu}}{q^{2}-m^{2}_{\pi}}\, \psibar{u}_{f}\,
          \frac{(\slashed{q}+\slashed{k_{\pi}})}{2} \, u_{i}  \notag
          \\[5pt]
&&{}+\frac{\epsilon^{i}_{\,jk}}{f_{\pi}}\, \bra {B} \uhalftau{k}\ket{A}\,
          4\kappa_{\pi}\,\psibar{u}_{f} \left(\frac{\usigma{\mn}ik_{\pi \nu}}{2M}
          +\frac{q^{\mu}}{q^{2}-m^{2}_{\pi}}\,
          \frac{\usigma{\ab}i k_{\pi\alpha}q_{\beta}}{2M} \right)u_{i} \notag
          \\[5pt]
&&{}+\frac{\delta_{j}^{\,i}}{f_{\pi}}\,
          \delta_{B}^{\,A}\, (-4 i\beta_{\pi})
          \frac{1}{M}\left(k_{\pi}^{\mu}-\frac{ q\cdot k_{\pi}\, q^{\mu}}{q^{2}-m_{\pi}^{2}}\right)
          \psibar{u}_{f}u_{i} \notag
          \\[5pt]
&&{}+\frac{\delta_{j}^{\,i}}{f_{\pi}}\, \delta_{B}^{\,A}\,
          \frac{-i\kappa_{1}}{4}\, \frac{1}{M^{2}}\,
          \psibar{u}_{f}\left(q_{\nu}(p_{f}+p_{i})^{\{\nu}\ugamma{\mu\}}
          -\frac{q \cdot (p_{f}+p_{i})\,
          q^{\mu}}{q^{2}-m_{\pi}^{2}}\,
          (\slashed{q}+\slashed{k_{\pi}})\right) u_{i} \, .
          \label{eqn:accontact}
\end{eqnarray}
\end{widetext}
Here, $\Gamma_{A\pi }^{\mu}(q,k_{\pi})$ is given in
Eq.~(\ref{eqn:NNpionaxialcurrentwithff}). The terms in the first row lead to 
$\nu\geqslant1$ contributions. The contributions due to
$\kappa_{\pi}$, $\beta_{\pi}$, and $\kappa_{1}$ are at $\nu=2$. 
We use values fitted in \cite{tang97simple} for these couplings. 
In the 
last row, ${A}^{\{\mu}{B}^{\nu\}}={A}^{\mu} {B}^{\nu}+{A}^{\nu} {B}^{\mu}$.

For the baryon current, diagrams (e) and (f) do not contribute: $\langle J_{B}^{\mu}\rangle_{\pi}=0.$

For the NC photon production matrix element we find
\begin{widetext}
\begin{eqnarray}
\langle J_{NC}^{\mu} \rangle_{\gamma} &=&
          \delta_{B}^{A}\, \frac{-iec_{1}}{M^{2}}\,
          \epsilon^{\mn\alpha\beta}\,
          \psibar{u}_{f}\dgamma{\nu}k_{\alpha}\epsilon^{\ast}_{\beta}(k) u_{i} 
           + \delta_{B}^{A}\, \frac{-iec_{1}
          q^{\mu}}{M^{2}(q^{2}-m^{2}_{\pi})}\, \epsilon^{\lambda \nu
          \alpha\beta}\, \psibar{u}_{f}\dgamma{\lambda} q_{\nu} k_{\alpha}
          \epsilon^{\ast}_{\beta}(k) u_{i} \notag
          \\[5pt]
&&{}+\left(\uhalftau{0}\right)_{B}^{\;A}\, \frac{-ie e_{1}}{2 M^{2}}\,
          \epsilon^{\mn\alpha\beta}\, \psibar{u}_{f}\dgamma{\nu}k_{\alpha}\epsilon^{\ast}_{\beta}(k)
          u_{i} \notag
          +\left(\uhalftau{0}\right)_{B}^{\;A}\, \frac{-ie e_{1} q^{\mu}}{2
          M^{2}(q^{2}-m^{2}_{\pi})}\, \epsilon^{\lambda \nu
          \alpha\beta}\,
          \psibar{u}_{f}\dgamma{\lambda} q_{\nu} k_{\alpha}
          \epsilon^{\ast}_{\beta}(k) u_{i} \ .
          \label{eqn:nccontact}
\end{eqnarray}
\end{widetext}
Here $\nu=3$; for $\nu < 3$, there are no contact vertices contributing in 
this channel. By power counting, we expect that at low
energy, these terms can be neglected compared to the $\nu=1$ terms.
However according to Ref.~\cite{RHill09}, these
terms may play an important role in coherent photon production.
Meanwhile, it is claimed in Ref.~\cite{RHill09} that the origin of 
these contact vertices is the anomalous interactions
of the $\omega$ and $\rho$. But they can
also be induced by the off-shell terms in the $\Delta$ Lagrangian. 
Moreover, we can construct meson-dominance terms by using the interaction terms in the 
last row of Eq.~(\ref{eqn:c1rhoe1rho}) and photon-meson coupling in 
Eq.~(\ref{eqn:photonmeson}), 
which leads to different off-shell behavior of the vertex compared to that of 
the anomaly term. 

\section{Results}
\label{sec:res}

In this section, after introducing the kinematics, we discuss
our results for CC and NC pion production, and also NC photon
production, and compare them with available data whenever possible.

\subsection{Kinematics}
\label{subsec:kinematics}

\begin{figure}[!ht]
\centering
 \includegraphics[scale=0.65]{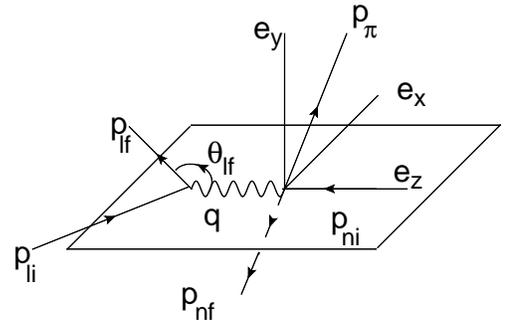}
 \caption{The configuration in the isobaric frame.}
 \label{fig:isobaricconfig}
\end{figure}

Fig.~\ref{fig:isobaricconfig} shows the configuration in the
isobaric frame, i.e., the center-of-mass frame of the final nucleon and pion.
The momenta are measured in this frame, except those labeled as
$p^{L}$, which are measured in the Lab frame with the initial
nucleon being static. Detailed analysis of the kinematics is
given in Appendix~\ref{app:kinmatics}. The expression for the total
cross section is 
%
\begin{align}
\sigma 
&=\int \frac{\overline{\vert M \vert^{2}}}{32 M_{n}}\,
          \frac{1}{(2\pi)^{5}}\,  \frac{\modular{p}{\pi}}{E_{\pi}+E_{nf}}
          \frac{\vert \vec{p}_{lf}^{\;L}\vert }{\vert \vec{p}_{li}^{\;L}\vert}\,
          d\Omega_{\pi} dE_{lf}^{L} d\Omega_{lf}^{L}  \notag
          \\[5pt]
&=\int \frac{\overline{\vert M \vert^{2}}}{64M_{n}^{2}}\,
          \frac{1}{(2\pi)^{5}}\,
          \frac{\modular{p}{\pi}}{E_{\pi}+E_{nf}}\,
          \frac{\pi}{\vert \vec{p}_{li}^{\;L}\vert E_{li}^{L}}\, d\Omega_{\pi}
          dM_{\pi n}^{2} d Q^{2} \ , 
\end{align}
%
where $\overline{\vert M
\vert^{2}}$ is the average of total interaction amplitude
squared.
Based on the equations in Appendix~\ref{app:kinmatics}, we can make
the following estimates.
For CC pion production,
when $E_{\nu}^{L}=0.4 \, (0.5) \, \mathrm{GeV}$, $(M_{\pi n})_{max}
      \approxeq 1.17 \, (1.24) \, \mathrm{GeV}, Q^{2}_{max} \approxeq 0.2 \, (0.3)\,
      \mathrm{GeV}^{2}$.
We can see that above $E_{\nu}^{L}=0.4 \,\mathrm{GeV}$, the
interaction begins to be dominated by the $\Delta$ resonance.
However, when $E_{\nu}^{L}=0.75 \, \mathrm{GeV}$, $(M_{\pi n})_{max}
\approxeq 1.4 \, \mathrm{GeV}$, and higher resonances, for example
$P_{11}(1440)$, may play a role. The exception is 
 $\nu_{\mu} + p \longrightarrow \mu^{-}+ p + \pi^{+}$: only $I=
3/2$ can contribute, and the next resonance in this channel is the
$\Delta(1600)$, which is accessible only when $E_{\nu}^{L} \geqslant
1.8\, \mathrm{GeV}$.
For NC pion production and photon production 
($E_{\gamma}^{L} \geqslant 0.2\, \mathrm{GeV} $),
when $E_{\nu}^{L}=0.3 \, (0.5) \, \mathrm{GeV}$, $(M_{\pi n})_{max}
      \approxeq 1.2 \, (1.35) \, \mathrm{GeV}, Q^{2}_{max} \approxeq 0.1 \,
      (0.3) \, \mathrm{GeV}^{2}$.
Here above $E_{\nu}^{L}=0.3\, \mathrm{GeV}$, the
interaction begins to be dominated by the $\Delta$. However, when
$E_{\nu}^{L}=0.6\, \mathrm{GeV}$, $(M_{\pi n})_{max} \approxeq 1.4
\, \mathrm{GeV}$, and higher resonances may play a role.

From this analysis, we expect our EFT to be valid
at $E_{\nu}^{L}\leqslant 0.5 \, \mathrm{GeV}$, since only the
$\Delta$ resonance can be excited, and $Q^{2} \leqslant 0.3 \,
\mathrm{GeV}^{2}$ where meson dominance works for various
currents' form factors \cite{BDS10}. To go beyond this energy regime
when we show our results, we require $M_{\pi n} \leqslant 1.4\,
\mathrm{GeV}$ and use phenomenological form factors
that work when $Q^{2} \geqslant 0.3 \, \mathrm{GeV}^{2}$.

\subsection{CC pion production}

\begin{figure}[!ht]
\centering
\includegraphics[scale=0.58,angle=-90]
{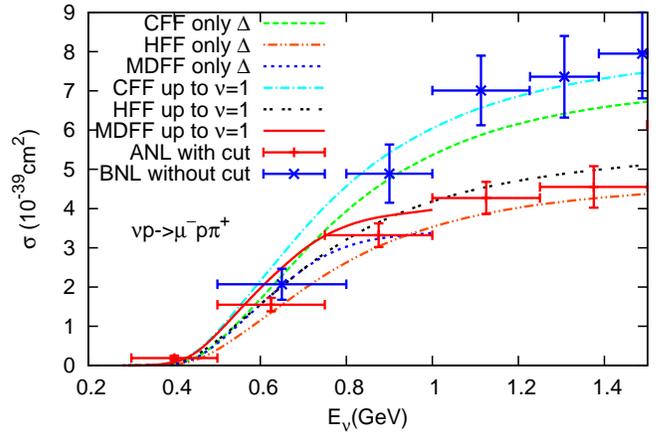}
\caption{(Color online). Total cross section for $\nu_{\mu}+p\longrightarrow
\mu^{-}+p+\pi^{+}$. ``Only $\Delta$'' indicates that only diagrams
with $\Delta$ (both $s$ and $u$ channels) are included. The ``up to
$\nu=1$'' category includes all the diagrams at leading order. The CFF
calculations are done with one of the conventional form
factors \cite{GRACZYK09}.
The HFF calculations make use of form factor used in \cite{HERNANDEZ07}
with the reduced $C_{5}^{A}(0)$. The MDFF calculations are based on the
EFT Lagrangian with meson dominance. In the ANL data, $M_{\pi n}
\leqslant$ 1.4 $\mathrm{GeV}$ is applied, while no such cut is
applied in the BNL data. For all calculations, $ M_{\pi n}
\leqslant$ 1.4 $\mathrm{GeV}$ is applied. } \label{Fig:pppion+}
\end{figure}

\begin{figure}[!ht]
\centering
\includegraphics[scale=0.58,angle=-90]
{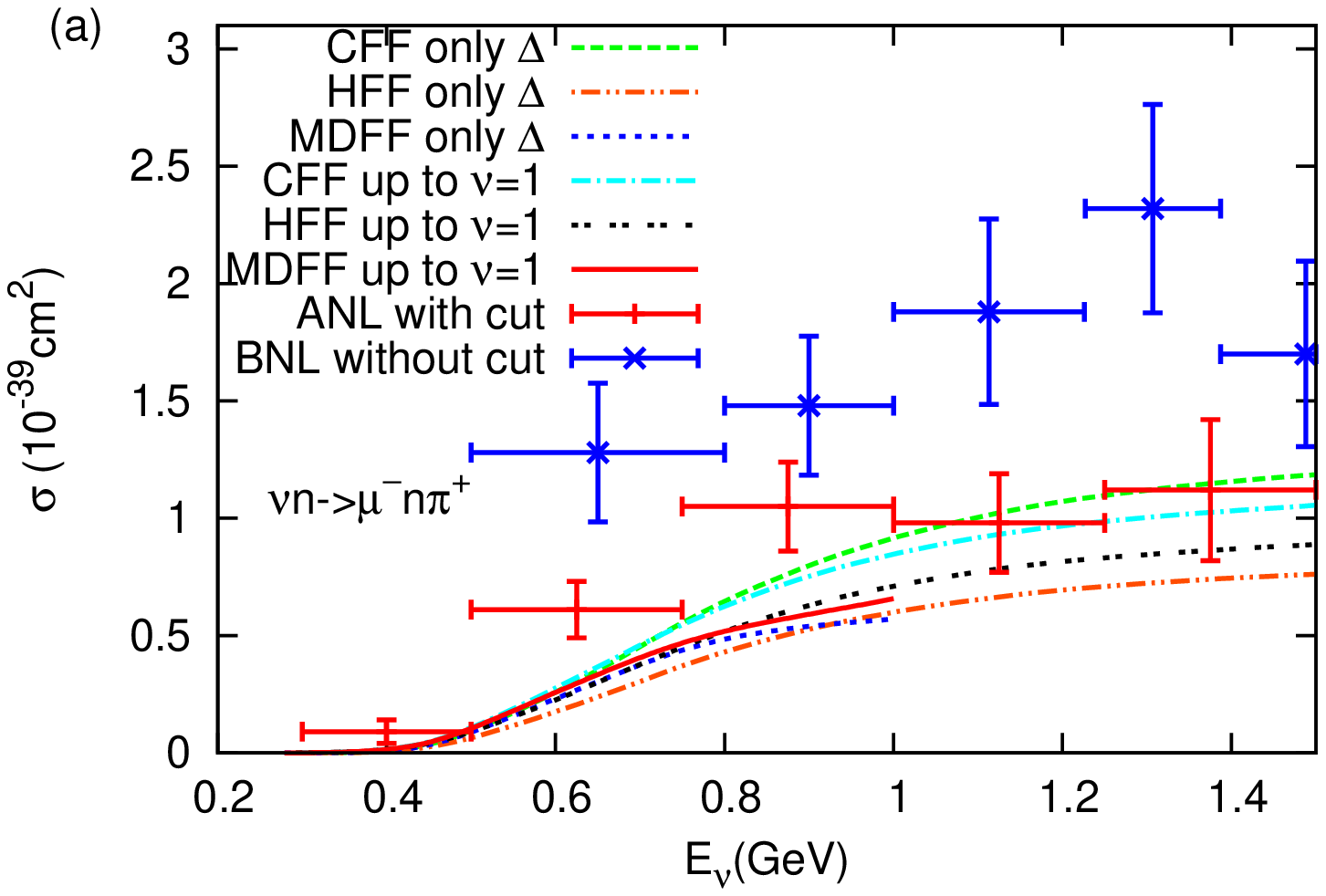}
\includegraphics[scale=0.58,angle=-90]
{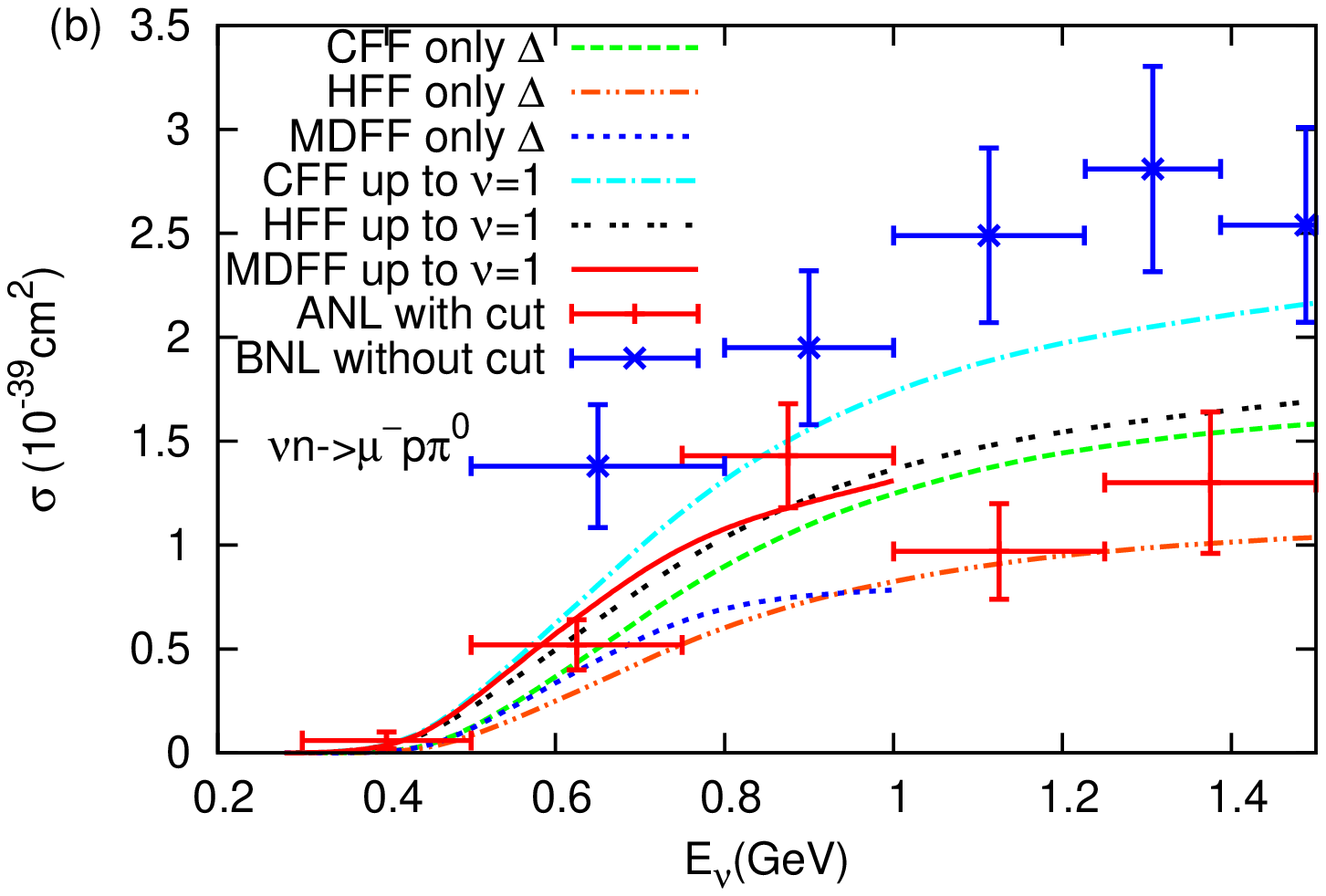}\caption{(Color online) Total cross section for (a) $\nu_{\mu}+n\longrightarrow
\mu^{-}+n+\pi^{+}$ and (b) $\nu_{\mu}+n\longrightarrow
\mu^{-}+p+\pi^{0}$. In the ANL data, $M_{\pi n}
\leqslant$ 1.4 $\mathrm{GeV}$ is applied, while no such cut is
applied in the BNL data. The curves are defined as in
Fig.~\ref{Fig:pppion+}.} \label{Fig:nnpion+0}
\end{figure}

In this section, we compare calculated cross sections of CC pion production 
with ANL~\cite{RADECKY82} and BNL~\cite{KITAGAKI86} measurements. In
both experiments, the targets are hydrogen and deuterium. [All the
other experiments use much heavier nuclear targets in (anti)neutrino
scattering, and to explain this, we must examine many-body effects.]
The beam is composed of muon neutrinos, the average energy of which is $1$ and $1.6\, \mathrm{GeV}$ for ANL and BNL respectively. 
In the ANL data, there is a cut on the invariant mass
of the pion and final nucleon system: $M_{\pi n} \leqslant 1.4\,
\mathrm{GeV}$; no such cut is applied in the BNL data. Based on the
previous phase-space analysis, this cut clearly reduces the
number of events when $E_{\nu}$ is above $\thicksim0.5\,
\mathrm{GeV}$. This can be seen by comparing the two data sets in three 
different channels shown in Figs.~\ref{Fig:pppion+} and \ref{Fig:nnpion+0}:
the ANL data lie systemically below the BNL data.
Since the data stretch above $0.5$ GeV, in Figs.~\ref{Fig:pppion+} and \ref{Fig:nnpion+0}, 
we show the ``CFF'' results [using the conventional form factor in \cite{GRACZYK09}]
and the ``HFF'' results [using the form factor in \cite{HERNANDEZ07} 
with the reduced $C_{5}^{A}(0)$], with the $M_{\pi n}$ constraint applied. 
In these calculations, $F^{md}$, $G^{md}$, $c_{\Delta}$, and $d_{\Delta}$ 
are substituted by the form factors in the literature.
The results of our framework, i.e. using the meson-dominance 
form factor born out of the Lagrangian, are shown as ``MDFF'' calculations, 
and these are extrapolated beyond 0.5 GeV limit also. 
The extrapolations of both CFF and MDFF calculations enable
us first to compare our result with similar calculations in \cite{HERNANDEZ07}, \footnote{The calculation in \cite{HERNANDEZ07} without reduction of $C_{5}^{A}(0)$ 
should be close to the CFF calculation \cite{GRACZYK09}, although the details of the form factors 
are different.}
and second to see how meson-dominance form factors fail at higher energy.
By comparing CFF with MDFF calculations, we can see in the MDFF 
calculation that the meson-dominance form factors are inadequate for reproducing 
the conventional form factors above $E_{\nu}=0.5\, \mathrm{GeV}$ 
(although it seems that MDFF results are closer to the data). 
Hence in the following Fig.~\ref{Fig:cclow}, we only show the MDFF results with $E_{\nu}
\leqslant 0.5 \thicksim 0.6\, \mathrm{GeV}$, for which $M_{\pi n}
\leqslant 1.4\, \mathrm{GeV}$ holds automatically. Since we believe
the EFT is applicable in this low-energy regime, in these plots,
we show results including Feynman diagrams up to order $\nu=1$ and
$\nu=2$.

In Fig.~\ref{Fig:pppion+}, we show the data and calculations for
$\nu_{\mu}+p\longrightarrow \mu^{-}+p+\pi^{+}$. 
As mentioned above, in the ``CFF only $\Delta$'' calculation, we make use of one set of 
conventional form factors and include 
the Feynman diagrams with the $\Delta$ in both $s$ and
$u$ channels. In the ``CFF up to $\nu=1$''calculation, we use the same form factors 
and include all the Feynman diagrams up to leading order. 
These two calculations are quite similar to
those done in Ref.~\cite{HERNANDEZ07} \emph{without} reducing $C_{5}^{A}$.
Indeed, our results are
consistent with theirs.
(In Ref.~\cite{HERNANDEZ07}, only the $s$ channel contribution is
included in the calculation with ``only $\Delta$.'') Next,
we show two different HFF calculations: one with only $\Delta$ (in the $s$
and $u$ channels) and the other with all the diagrams up to $\nu=1$. 
Finally, we also show two MDFF calculations up to different order, so that
we can compare the MDFF approach with the CFF approach.

First, we can see that both CFF and MDFF calculations with only $\Delta$
diagrams are consistent with the data at $E_{\nu} \leqslant 0.5 \,
\mathrm{GeV}$. Introducing other diagrams up to order $\nu=1$ is
still allowed by the data at low energy, although they indeed
increase the cross section noticeably. Second, 
in Ref.~\cite{HERNANDEZ07}, a reduced $C_{5}^{A}(0)$ is
introduced, primarily to reduce the calculated cross sections above
$E_{\nu}=1\, \mathrm{GeV}$, which can be seen by comparing CFF
calculations with HFF calculations. However, since we are only concerned
with the $E_{\nu} \leqslant 0.5\, \mathrm{GeV}$ region, in which we
see satisfactory agreement between our calculations and the data, we
will keep the $C_{5}^{A}(0)$ fitted from the $\Delta$'s free width. 
Furthermore, in the original
spectrum-averaged $d\sigma/dQ^{2}$ data of ANL~\cite{RADECKY82}, the
contributions from $E_{\nu} \leqslant 0.5\, \mathrm{GeV}$ neutrinos
are excluded, so comparing calculations with data at low energy is
not feasible at this stage, and we will not show our
$d\sigma/dQ^{2}$ here.

In Fig.~\ref{Fig:nnpion+0}, we show the data
and calculations for $\nu_{\mu}+n\longrightarrow \mu^{-}+n+\pi^{+}$
and $\nu_{\mu}+n\longrightarrow \mu^{-}+p+\pi^{0}$. We can see that
the situations in these two processes are quite similar to that
in Fig.~\ref{Fig:pppion+}: the results of the CFF and MDFF
approaches are consistent with the data at low energy. Again the
differences between the two approaches with the same diagrams begin
to show up when the neutrino energy goes beyond $0.5\,
\mathrm{GeV}$. Although the pion production is still dominated by
the $\Delta$, if we compare cross sections (from the same calculation) 
in Figs.~\ref{Fig:pppion+} and \ref{Fig:nnpion+0},
we see that other diagrams introduce significant contributions and 
violate the naive estimate of the ratio of the three channels'
cross sections based on isospin symmetry and $\Delta$ dominance.
Moreover, the reduction of $C_{5}^{A}$ significantly reduces the cross section in these two channels if we compare the two
HFF calculations with the corresponding CFF calculations.

\begin{figure*}[!ht]
\centering
\includegraphics[scale=0.71,angle=-90]
{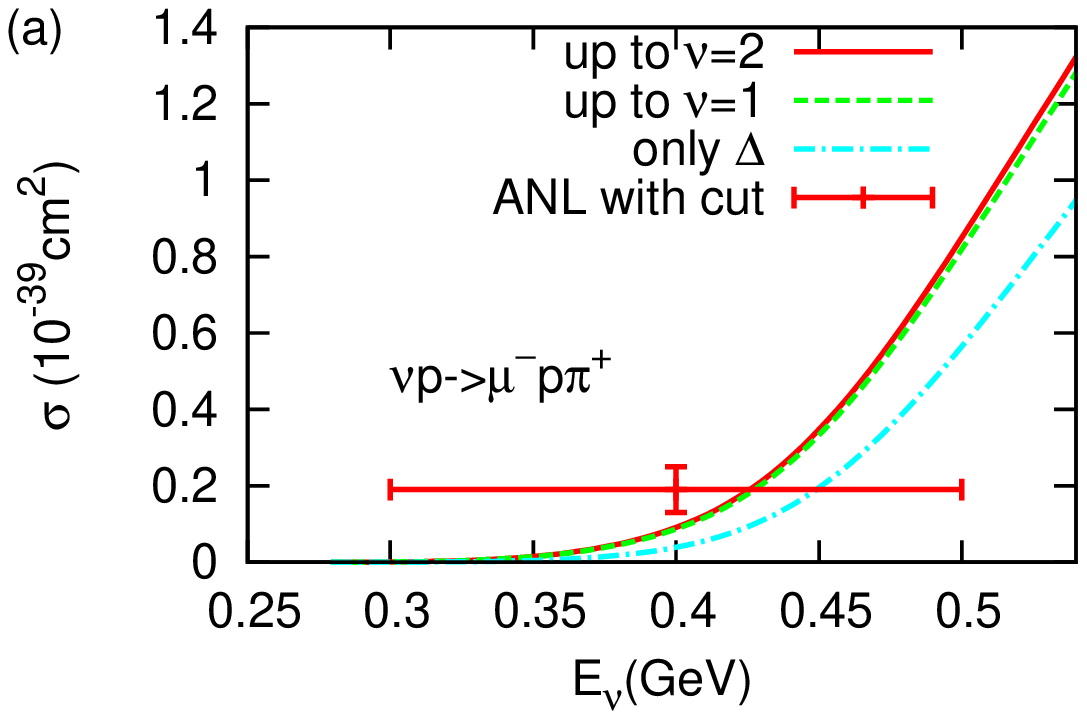}
\includegraphics[scale=0.71,angle=-90]
{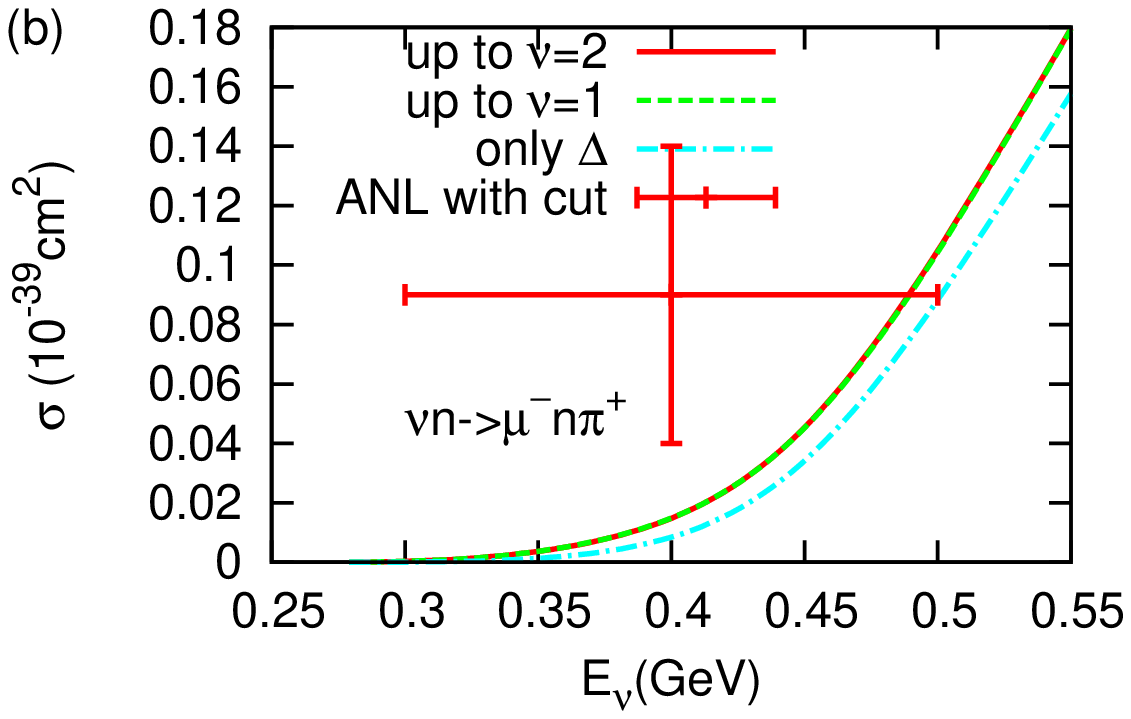}
\includegraphics[scale=0.71,angle=-90]
{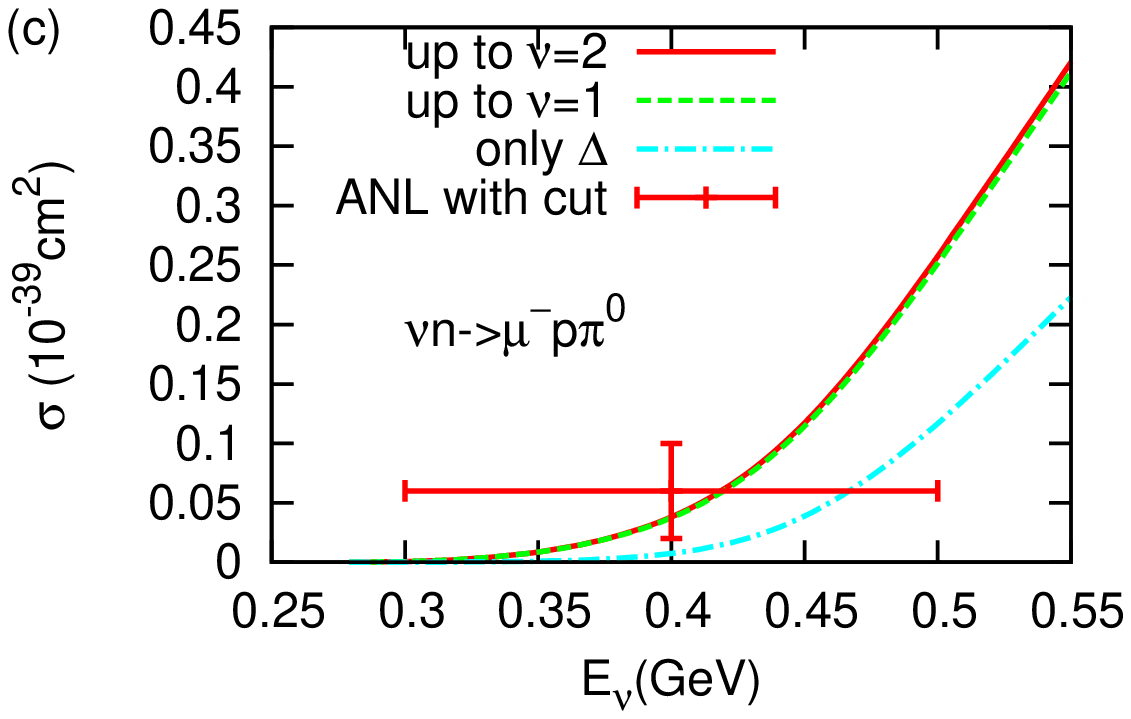}
\includegraphics[scale=0.715,angle=-90]
{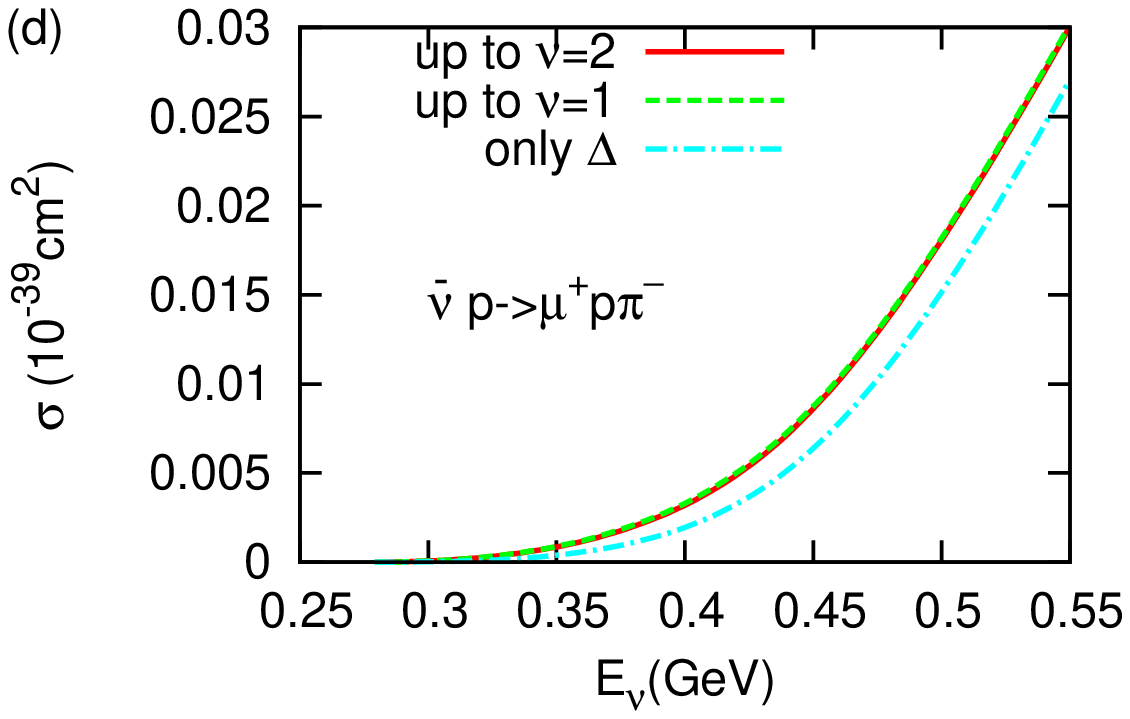}
\includegraphics[scale=0.71,angle=-90]
{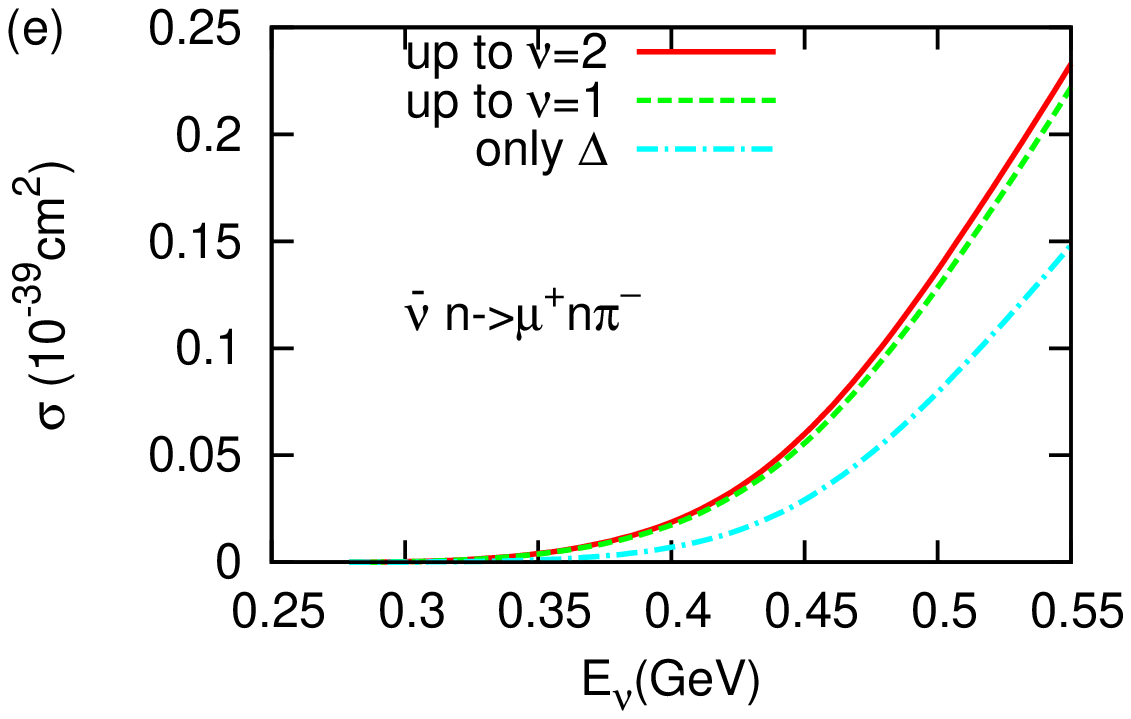}
\includegraphics[scale=0.71,angle=-90]
{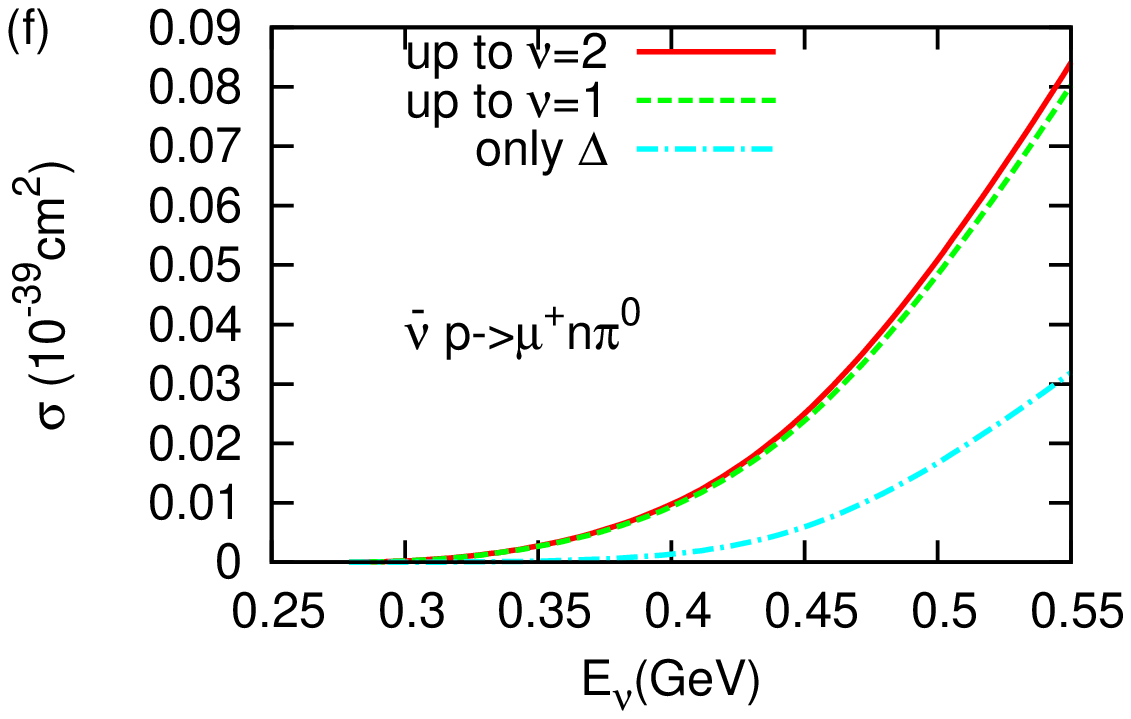}
\caption{(Color online) Total cross section for CC
pion production due to neutrino and antineutrino scattering off nucleons.
 Here ``only $\Delta$'' indicates that only diagrams
with $\Delta$ (both $s$ and $u$ channels) are included, ``up to
$\nu=1$'' includes all the diagrams at leading order, and ``up to $\nu=2$''
includes higher-order contact terms, whose couplings are from
\protect{Ref.~\cite{tang97simple}}. In the ANL data, $M_{\pi n}
\leqslant$ 1.4 $\mathrm{GeV}$. For calculations, $M_{\pi n}
\leqslant$ 1.4 $\mathrm{GeV}$ is applied. } \label{Fig:cclow}
\end{figure*}

In Fig.~\ref{Fig:cclow}, we begin to investigate the convergence
of our calculations in different channels in neutrino and
antineutrino scattering. We show the MDFF calculations based on
our EFT Lagrangian up to different orders. We see that the power
counting makes sense systematically in different channels: including
$N$ intermediate state and contact terms up to $\nu=1$ changes the ``only $\Delta$''
calculation non-negligibly. Far below resonance, 
the $\Delta$ contribution is less important compared to that in other
diagrams, and it begins to dominate around $ 0.4 \, \mathrm{GeV}$.
This is consistent with the power counting discussed in
Sec.~\ref{sec:diag}. Moreover, the $\nu=2$ terms do not
change the ``up to $\nu=1$'' results significantly. 
All the calculations of neutrino scattering are
consistent with the limited data from ANL. We can see that the cross
section for antineutrino scattering is generally smaller than that
of neutrino scattering, due to the relative sign chosen between
$V^{i\mu}$ and $A^{i\mu}$ in the Feynman diagrams having $\Delta$. The
sign between $V^{i\mu}$ and $A^{i\mu}$ in other diagrams is well
defined in our Lagrangian. The relative sign between
$\Delta$'s contribution and other diagrams' is also well
determined by the relation between $h_{A}$ and $C_{5}^{A}$ 
in Eq.~(\ref{eqn:axialtransitionff3}), 
although it has been investigated phenomenologically
in Ref.~\cite{HERNANDEZ07}.

\subsection{NC pion production}

\begin{figure*}[!ht]
\centering
\includegraphics[scale=0.71,angle=-90]
{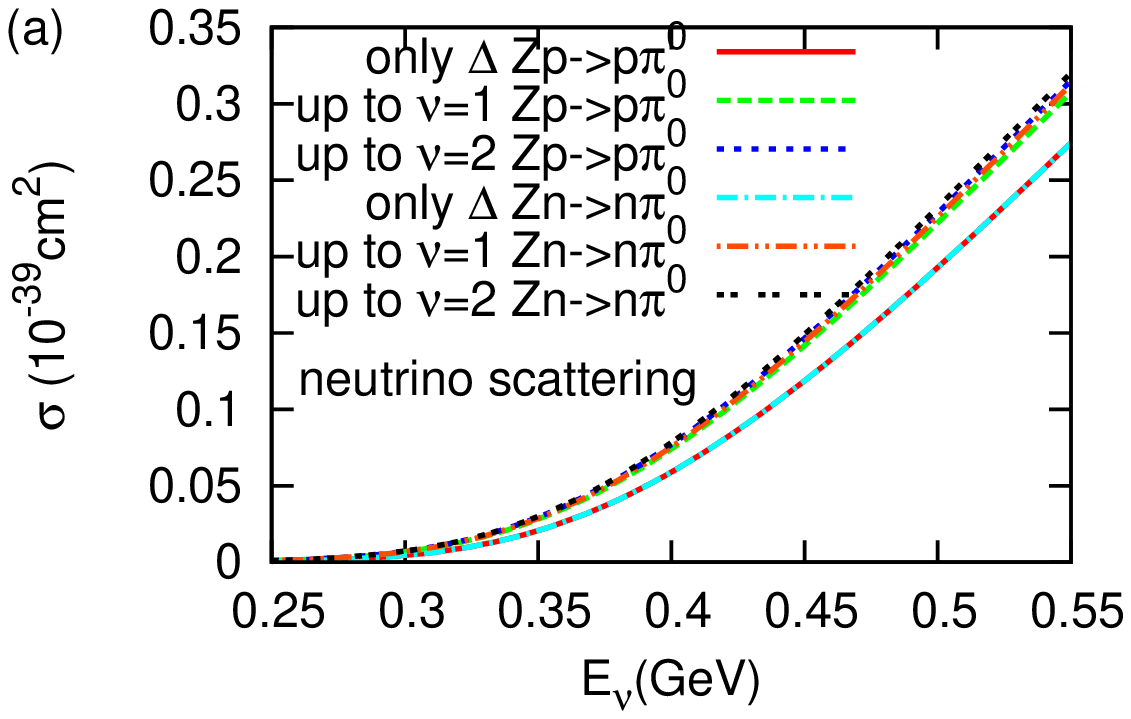}
\includegraphics[scale=0.71,angle=-90]
{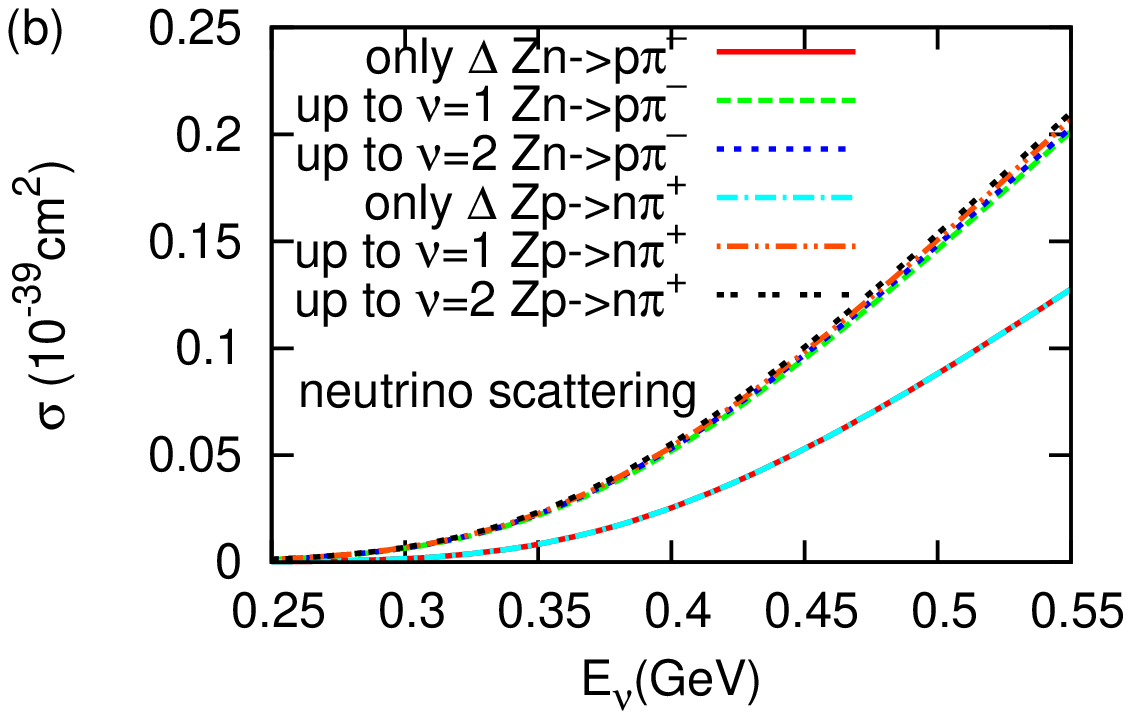}
\includegraphics[scale=0.71,angle=-90]
{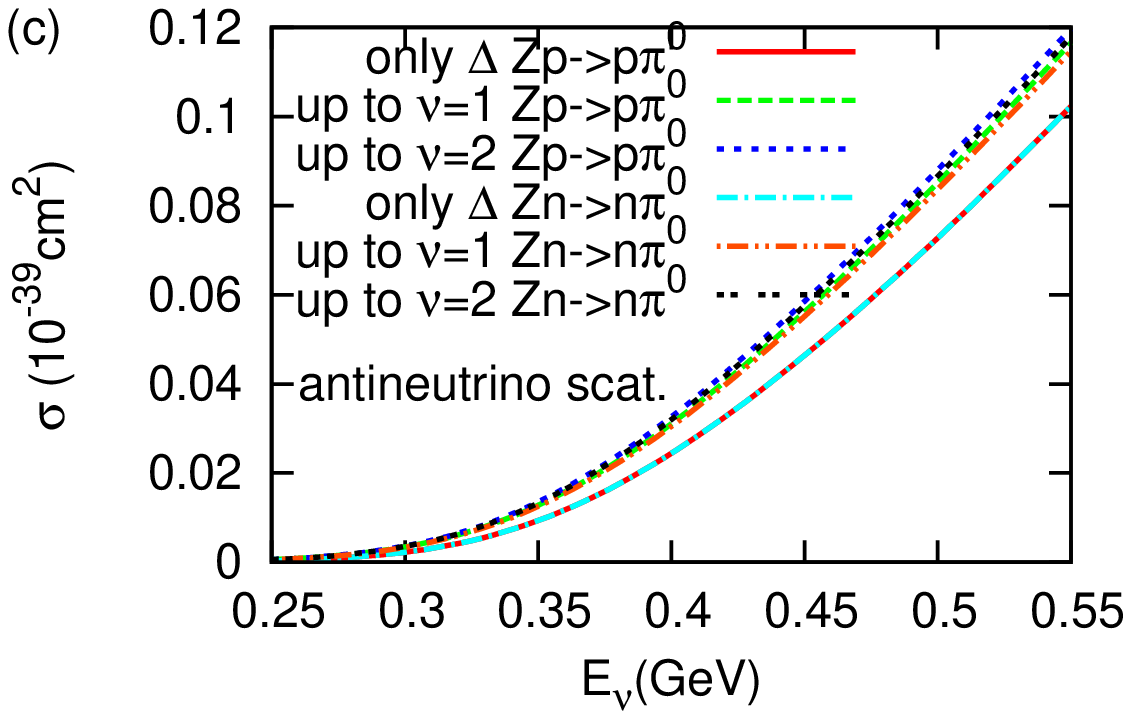}
\includegraphics[scale=0.71,angle=-90]
{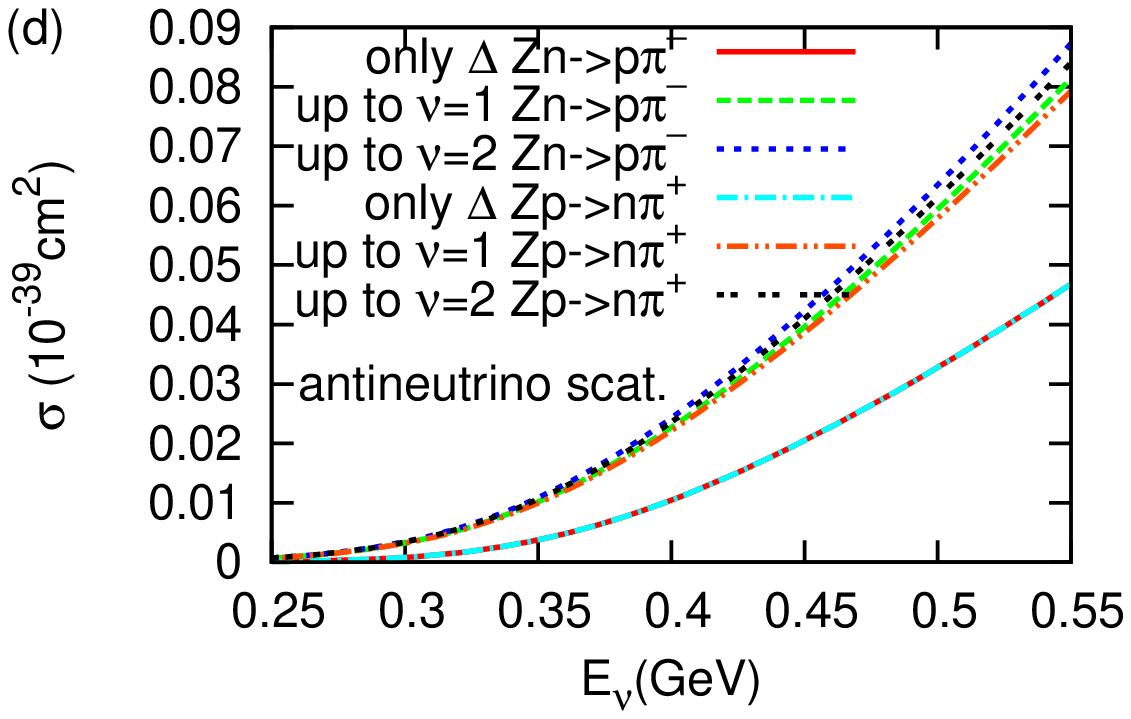}
\caption{(Color online). Total cross section for NC $\pi$ production due to
neutrino and antineutrino scattering off nucleons. The curves are defined as in
Fig.~\ref{Fig:cclow}, and the channels are also indicated. }
\label{Fig:ncpion}
\end{figure*}
In this section, we discuss the results for NC pion production in
(anti)neutrino scattering. In Fig.~\ref{Fig:ncpion}, 
the results in the MDFF approach are shown for calculations including
diagrams of different orders. The channels are explained in each plot.
Since all of the available data for NC pion production are 
spectrum-averaged, and neutrinos with $E_{\nu} \leqslant 0.5\, \mathrm{GeV}$
have a small weight in such analyses, we do not
compare our results with data. 
Here we focus on the convergence of our calculations;
introducing the $\nu=2$ terms does not change the total cross
section significantly. However, we also see the violation of isospin
symmetry in the ``up to $\nu=1$'' and ``up to $\nu=2$'' calculations in
each plot, if we compare each pair of channels in 
Fig.~\ref{Fig:ncpion}. In principle, if
there is no baryon current contribution in NC production, we should
see that the two channels in each plot yield the same results. 
For example, isospin symmetry implies
$\bra{p,\pi^{0}} V^{0\mu}, A^{0\mu} \ket{p} =
          \bra{n,\pi^{0}} V^{0\mu}, A^{0\mu} \ket{n} $
and
$\bra{p,\pi^{0}} J_{B}^{\mu} \ket{p} = -\bra{n,\pi^{0}}
          J_{B}^{\mu} \ket{n} $.
So with ``only $\Delta$,'' we can not see the difference between the
two cross sections, since the (isoscalar) baryon current cannot
induce transitions from $N$ to $\Delta$.  After introducing
nonresonant diagrams, we would expect them to be
different, as confirmed in the first plot in Fig.~\ref{Fig:ncpion} for example. 
This analysis can be applied to 
other channels, and we clearly see the nonresonant contributions. 

\subsection{NC photon production} \label{subsec:ncphotonprod}

\begin{figure*}[!ht]
\centering
\includegraphics[scale=0.71,angle=-90]
{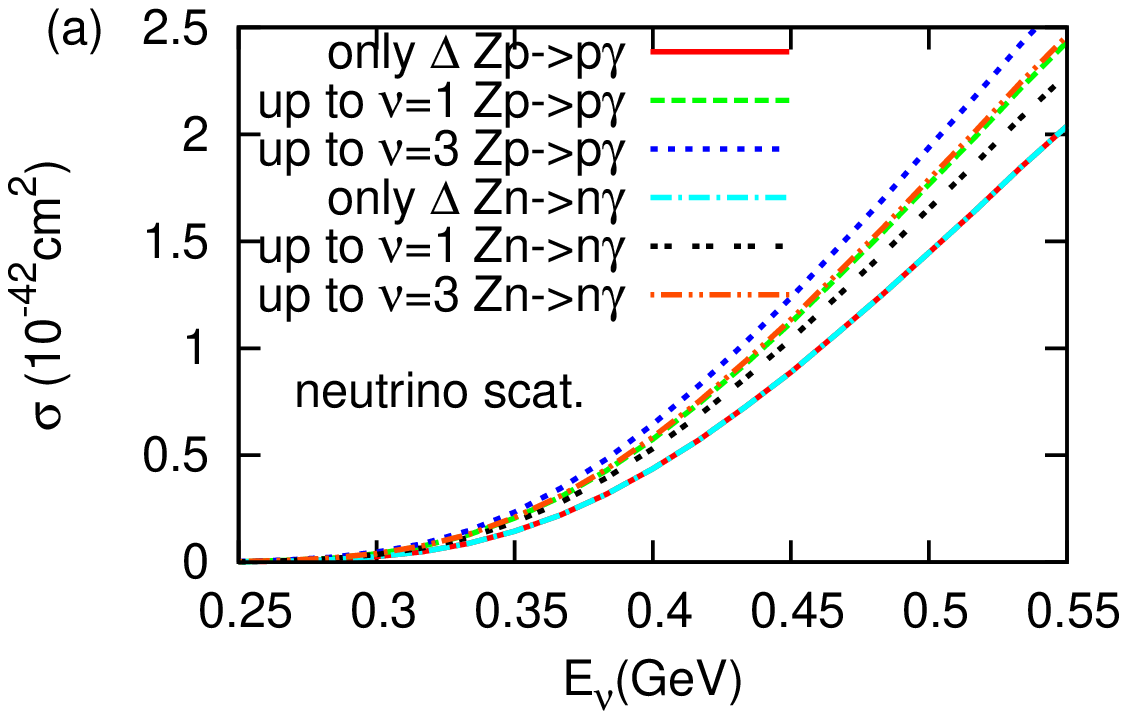}
\includegraphics[scale=0.71,angle=-90]
{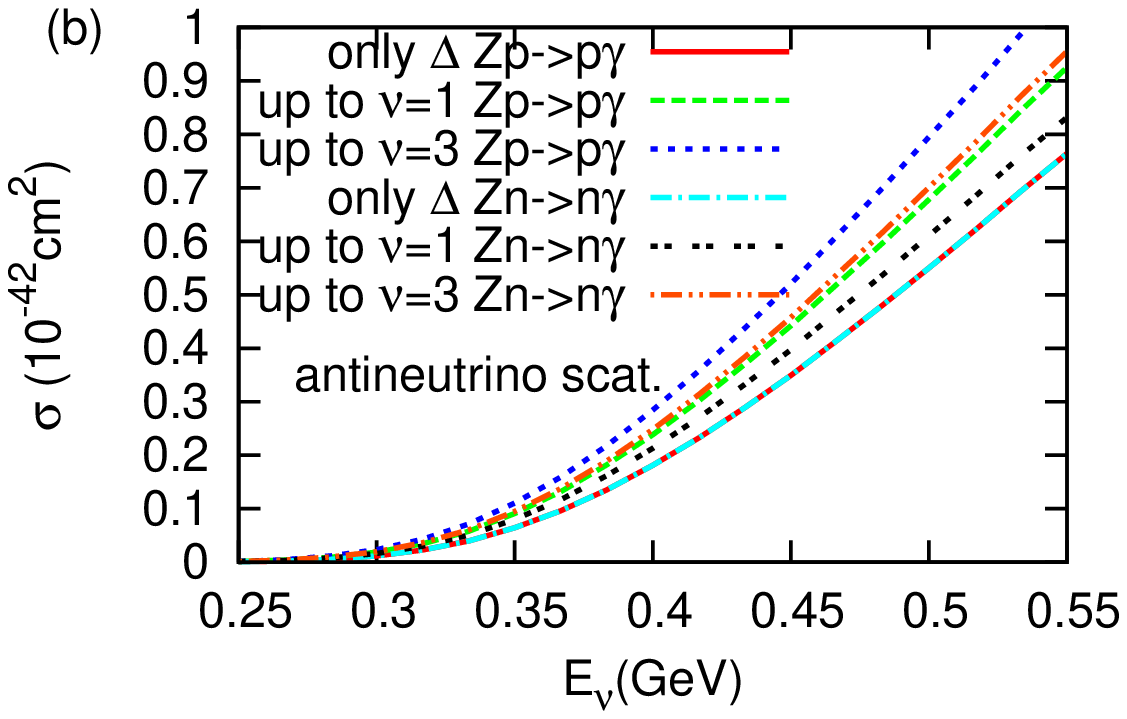}
\caption{(Color online) Total cross section for NC photon production due to
neutrino and antineutrino scattering off nucleons. Here ``only $\Delta$'' indicates that only diagrams
with $\Delta$ (both $s$ and $u$ channels) are included, ``up to
$\nu=1$'' includes all the diagrams at leading order, and ``up to $\nu=3$''
includes higher-order diagrams. }
\label{Fig:ncphoton}
\end{figure*}

In this section we focus on NC photon production. The results 
are shown in Fig.~\ref{Fig:ncphoton}. Besides NC
$\pi^{0}$ production, this process is another important background
in neutrino experiments. One important difference between NC photon
production and CC and NC pion production, is that all of the $\nu=2$
terms do not contribute in this process. Therefore, we include the
two $\nu=3$ terms in NC photon production, namely the $e_{1}$ and
$c_{1}$ couplings in Eq.~(\ref{eqn:nccontact}), besides terms due to the form factors. 
 Moreover, these
two couplings are singled out in Ref.~\cite{RHill09} as the
low-energy manifestations of anomalous interactions involving 
$\rho$ and $\omega$, and they 
are believed to give important contributions in coherent photon
production from nuclei. Here we also investigate the consequences of
these two couplings. We emphasize that from the EFT perspective, the
only way to determine these two couplings is by comparing the final
theoretical result with data, rather than by calculating them from
anomalous interactions, which are not necessarily the only high-energy
physics contributing to these two operators. For example, as we
discussed before, an off-shell coupling
between $N$, $\pi$, and $\Delta$ can introduce the same matrix
element as the $c_{1}$ term. Changing the
off-shell couplings would also change the contact term to
make the theory independent of the choice of off-shell couplings.
Nevertheless, to perform concrete calculations 
without precise information on the coupling strengths, we use the
values from Ref.~\cite{RHill09} ($c_{1}=1.5, e_{1}=0.8$).

We can see the convergence of our calculations in Figs.~\ref{Fig:ncphoton}. The two couplings
introduced in the ``up to $\nu=3$'' calculations increase the total
cross section in both channels for both neutrino and antineutrino
scattering, although the change is quite small. This constructive
behavior is consistent with the results in Ref.~\cite{RHill09}.

Naive power counting, however, does not give an accurate comparison
between the $\Delta$ contributions and the $N$ contributions at very 
low energy. 
First, the neutron does not have an electric
charge, so its current should appear at higher order
than the naive estimate would indicate. Second, for the proton,
due to the cancellation between the baryon current and the vector
current, the neutral current is mainly composed of the axial-vector
current, which reduces the strength of the neutral current. Because
of these two effects, the contributions of Compton-like diagrams are smaller 
than the power counting indicates.

\section{summary} \label{sec:sum}

Neutrinoproduction of photons and pions from nucleons and nuclei produce
important backgrounds in neutrino-oscillation experiments and therefore
must be understood quantitatively. In this work, we studied
the productions from free nucleons in a Lorentz-covariant, chirally
invariant, meson--baryon EFT. For (anti)neutrino energy around 
$0.5 \,\mathrm{GeV}$, the $\Delta$ resonance 
is important. We therefore included the $\Delta$
degrees of freedom explicitly in our EFT Lagrangian, in a manner
that is consistent with both Lorentz covariance and chiral symmetry.

It is well known that in a Lagrangian with a finite number of
interaction terms, including the $\Delta$ as a Rarita--Schwinger
field leads to inconsistencies for strong couplings, strong fields,
or large field variations. In a modern EFT with an infinite number
of interaction terms, however, these pathologies can be removed, if
we work at low energies with weak boson fields. This is clarified 
in our previous work \cite{SZbookchapter}.
Ambiguous and so-called off-shell couplings involving $\Delta$ 
have also been shown to be redundant in the modern EFT framework, 
because these couplings produce terms that can 
be absorbed into the contact terms in the EFT Lagrangian. 
Thus the $\Delta$ resonance can be introduced in our EFT 
Lagrangian in a consistent way. 

Because of the symmetries built into our Lagrangian, the vector 
and baryon currents are conserved and the axial-vector currents are partially conserved automatically, which is not true in some other approaches to
this problem (with special constraints among different form factors having  
to be introduced by hand to conserve vector current in other 
approaches). Needless to say, the conserved vector and baryon currents are
crucial for computing photon production. We discussed in detail how the 
meson-dominance mechanism works in our matrix element calculations, which is 
the key ingredient in current conservation. By using vector and
axial-vector transition currents that were calibrated in pion production 
at high energies, we found results for pion production at lower energies
($E_{\nu}^{\mathrm{Lab}} \leqslant 0.5$ GeV) that 
are consistent with the (limited) data. This is also true  
when vertices described by meson dominance were used. 
On the other hand, the couplings introduced to generate meson 
dominance are relevant in other problems. For example, the interactions 
in Eq.~(\ref{eqn:truelagdeltaNrho}) lead to a
proper description of the vector transition current at nonzero 
$Q^{2}$ and meanwhile it is relevant to two-body currents: 
suppose a photon is absorbed by one nucleon 
producing a $\Delta$ which then interacts with other nucleon through the 
interactions mentioned above. 
 
Moreover, we studied the convergence of our power-counting scheme 
at low energies (where the $\Delta$ needs to be counted differently in 
different energy regions) and found that next-to-leading-order tree-level corrections 
are small. This power counting scheme is different from the 
canonical one, because it can be used in nuclear many body problems. 
For example, the lowest order in this scheme is the mean-field approximation, if 
the calculation is done for the property of the nuclear ground states. 
The discussion on this can be found in \cite{MCINTIRE07,HU07,MCINTIRE08}. 
It is certainly interesting to see how the power counting that we have for 
scattering off nucleons works in the scattering off nuclei.  

Finally, we computed NC photon production and explored the power 
counting in this problem. The difference between the NC photon production 
and pion production is that, at $\nu=2$, no diagrams contribute in the photon 
case, while there are several in pion production. So we proceeded to 
include $\nu=3$ diagrams induced by two contact interactions, $c_{1}$ and 
$e_{1}$ terms. They have been studied in \cite{RHill09}, and they are believed to 
be the low energy manifestation of anomalous $\rho$ and $\omega$ interactions. 
We pointed out the existence of other sources including off-shell 
couplings of $\Delta$ and possible meson-dominance terms. 
Nevertheless, by using two coupling strengths calibrated to 
anomalous $\rho$ and $\omega$ interactions \cite{RHill09}, 
we found that, at least for a nucleon target, their contributions 
are very small, as expected based on power counting. 

We are currently using this QHD EFT framework to study the
electroweak response of the nuclear many-body system, so that we can
extend our results to photon and pion neutrinoproduction from
nuclei, which are the true targets in existing neutrino-oscillation
experiments.

\acknowledgments

XZ thanks Mikhail Gorshteyn for the careful reading of the manuscript. 
This work was supported in part by the Department of Energy under
Contract No.\ DE--FG02--87ER40365.

\appendix

\section{Chiral symmetry and electroweak interactions in QHD EFT} \label{app:chisym}
Details on chiral symmetry and electroweak interactions in QHD EFT can be found in \cite{SZbookchapter,XZThesis}. 
We introduce background fields including $\vbg^{\mu}\equiv
\vbg^{i\mu} \tau_{i}/{2}$ \ (isovector vector), $\vbg_{(s)}^{\mu}$ \
(isoscalar vector), and $\abg^{\mu}\equiv \abg^{i\mu} \tau_{i}/{2}$ \
(isovector axial-vector), where $i=x,y,z \ \text{or} \ +1,0,-1$. 
They couple to the corresponding currents in QCD.   
We define $r^{\mu}= \vbg^{\mu}+\abg^{\mu}$, and 
$l^{\mu}=\vbg^{\mu}-\abg^{\mu}$. 
Under $SU(2)_L \otimes SU(2)_R \otimes U(1)_B$ symmetry transformations, 
these fields should change in the following way: 
$l^{\mu} \to L \, l^{\mu} L^{\dagger} + i L \, \partial^{\mu} L^{\dagger} $ 
        \{$L=\exp [ -i\theta_{Li}(x)\,\frac{\tau^{i}}{2} ] $\}, 
$r^{\mu} \to R \, r^{\mu} R^{\dagger} + i R \, \partial^{\mu} R^{\dagger} $ 
       \{$ R=\exp [ -i\theta_{Ri}(x)\,\frac{\tau^{i}}{2} ] $\},
and $\vbg_{(s)}^{\mu} \to \vbg_{(s)}^{\mu} - \partial^{\mu} \theta $. 
Here, $\theta_{Li}(x)$, $\theta_{Ri}(x)$, and $\theta(x)$ are the rotation 
angles.  
We can construct field strength tensors: 
$f_{L\mu\nu}\equiv \partial_{\mu}l_{\nu}-\partial_{\nu}l_{\mu}
-i \left[l_{\mu} \, , \, l_{\nu}\right] \to L f_{L\mu\nu}L^{\dagger}$, 
and $f_{R\mu\nu}$ and $f_{s\mu\nu}$ are constructed in the same way.

Now we discuss nonlinear transformations of dynamical 
degrees freedom in our model:
\begin{eqnarray}
U&\equiv&\exp \left[ 2i\frac{\pi_{i}(x)}{f_{\pi}}\, t^{i} \right]
       \to  LUR^{\dagger} \ ,  \label{eqn:allchiral1} \\[5pt]
\xi&\equiv& \sqrt{U}= \exp \left[ i\frac{\pi_{i}}{f_{\pi}}\, t^{i}
       \right]
       \to L\xi h^{\dagger}=h \,\xi R^{\dagger} \ , \label{eqn:allchiral2} \\[5pt]
\widetilde v_{\mu}&\equiv& \frac{-i}{2} [\xi^{\dagger}(\partial_{\mu}
       -il_{\mu})\xi+\xi(\partial_{\mu}-ir_{\mu})\xi^{\dagger}]
       \equiv \widetilde v_{i\mu}t^{i} \notag \\[5pt]
      & \to & h \, \widetilde v_{\mu} h^{\dagger} -ih \, \partial_{\mu}h^{\dagger} \ ,
        \label{eqn:allchiral3} \\[5pt]
\widetilde a_{\mu}&\equiv& \frac{-i}{2} [\xi^{\dagger}(\partial_{\mu}
       -il_{\mu})\xi-\xi(\partial_{\mu}-ir_{\mu})\xi^{\dagger}]
       \equiv \widetilde a_{i\mu}t^{i} \notag \\[5pt] 
      & \to & h \, \widetilde a_{\mu} h^{\dagger} \ ,
       \label{eqn:allchiral4} 
\end{eqnarray}

\begin{eqnarray}      
\dcpartial{\mu}U&\equiv& \partial_{\mu} U -i l_{\mu} U +i U r_{\mu}
       \to L \, \dcpartial{\mu}UR^{\dagger} \ ,  \label{eqn:allchiral5}\\[5pt]
(\dcpartial{\mu}\psi)_{\alpha}&\equiv& (\partial_{\mu}+i\,\widetilde v_{\mu}
       -i\vbg_{(s)\mu}B)_{\alpha}^{\;\beta} \psi_{\beta} \notag \\[5pt]
      & \to & \exp \left[ -i\theta(x)B \right]
       h_{\alpha}^{\;\beta} (\dcpartial{\mu}\psi)_{\beta} \ , \label{eqn:allchiral6}
       \\[5pt]
\widetilde{v}_{\mu\nu}&\equiv& -i [\widetilde{a}_{\mu} \, , \,
       \widetilde{a}_{\nu}]
       \to h \, \widetilde{v}_{\mu\nu} h^{\dagger} \ ,  \label{eqn:allchiral7}
       \\[5pt]
F^{(\pm)}_{\mu\nu}&\equiv&\xi^{\dagger}f_{L\mu\nu}\,\xi
       \pm \xi f_{R\mu\nu}\, \xi^{\dagger}
       \to hF^{(\pm)}_{\mu\nu}h^{\dagger} \ ,  \label{eqn:allchiral8} \\[5pt]
\dcpartial{\lambda} F^{(\pm)}_{\mu\nu} &\equiv&
       \partial_{\lambda}F^{(\pm)}_{\mu\nu} + i [\widetilde{v}_{\lambda}
       \, , \, F^{(\pm)}_{\mu\nu}] \to h\,\dcpartial{\lambda}
F^{(\pm)}_{\mu\nu}h^{\dagger} \ . \label{eqn:allchiral10} 
\end{eqnarray}
In the preceding equations, $t^{i}$ are the generators of reducible
representations of $SU(2)$. The $f_{\pi}\approx 93$ MeV is the pion-decay constant.
We generically label non-Goldstone 
isospin multiplets including the nucleon, $\rho$ meson, and $\Delta$ by
$\psi_{\alpha}=\left( N_{A}, \rho_{i}, \Delta_{a} \right)_{\alpha}$.
$B$ is the baryon number of the particle. The transformations of
the isospin and chiral singlets $V_{\mu}$ and $\phi$ are trivial. 
The dual field tensors are defined as
$\psibar{F}^{\,(\pm)\,\mu\nu} \equiv \epsilon^{\mu\nu\alpha\beta}
F^{(\pm)}_{\alpha\beta}$, which have the same chiral transformations
as the ordinary field tensors. The objects shown here are the 
building blocks for constructing the Lagrangian. 

Electroweak interactions of quarks in the Standard Model 
\cite{IZ80,DGH92,SZbookchapter,XZThesis} determine
the form of the background fields in terms of the vector bosons
$W^{\pm}_{\mu}$, $Z_{\mu}$, and $A_{\mu}$:
\begin{eqnarray}
l_{\mu}&=&-e\, \frac{\tau^{0}}{2}\, A_{\mu}
       +\frac{g}{\cos\theta_{w}}\sin^{2}\theta_{w}\, \frac{\tau^{0}}{2}\, Z_{\mu}  
       -\frac{g}{\cos\theta_{w}}\frac{\tau^{0}}{2}\, Z_{\mu} \notag
       \label{eqn.lmubackground} \\[5pt]
&& {} -g V_{ud}\, \left( W^{+1}_{\mu}\, \frac{\tau_{+1}}{2}
       +W^{-1}_{\mu}\frac{\tau_{-1}}{2} \right) 
       \ , \\[5pt]
r_{\mu}&=&-e\, \frac{\tau^{0}}{2}\, A_{\mu}
       +\frac{g}{\cos\theta_{w}}\sin^{2}\theta_{w}\, \frac{\tau^{0}}{2}\, Z_{\mu} 
       \ , \label{eqn.rmubackground} \\[5pt]
\vbg_{(s)\mu}&=&-e\, \half\,
       A_{\mu}+\frac{g}{\cos\theta_{w}}\sin^{2}\theta_{w}\, \half\,
       Z_{\mu} \ , \label{eqn.vsmubackground}
\end{eqnarray}
where $g$ is the $SU(2)$ charge, $\theta_{w}$ is the weak mixing
angle, and $V_{ud}$ is the CKM matrix element corresponding
to $u$ and $d$ quark mixing.

If we define the interactions with background fields as  
\begin{eqnarray}
\lag_{\mathrm{ext}}&\equiv&\vbg_{i\mu}V^{i\mu}-\abg_{i\mu}A^{i\mu}+\vbg_{(s)\mu}J^{B\mu}
       \notag \\[5pt]
&=& J^{L}_{i\mu}\, l^{i\mu}+J^{R}_{i\mu}\, r^{i\mu}+\vbg_{(s)\mu}J^{B\mu} \ , 
\end{eqnarray}
define electroweak interactions as
\begin{align}
\lag_{I}&= -eJ^{EM}_{\mu}A^{\mu}
       -\frac{g}{\cos\theta_{w}}J^{NC}_{\mu}Z^{\mu} \notag \\[5pt]
      & -gV_{ud}\,J^{L}_{+1 \mu}W^{+1\mu}-gV_{ud}\, J^{L}_{ -1 \mu}W^{-1\mu}\ ,  
\end{align}
and use Eqs.~(\ref{eqn.lmubackground}) to
(\ref{eqn.vsmubackground}), we can see that 
\begin{eqnarray}
J^{L}_{i \mu} &\equiv& \half\, (V_{i\mu}+A_{i\mu}) \ , 
       \\[5pt]
J^{R}_{i \mu} &\equiv& \half\, (V_{i\mu}-A_{i\mu}) \ , 
       \\[5pt]
J^{EM}_{\mu}&=&V^{0}_{\mu}+\half\, J^{B}_{\mu} \ , 
       \\[5pt]
J^{NC}_{\mu}&=& J^{L0}_{\mu}-\sin^{2}\theta_{w}\, J^{EM}_{\mu} \ .
       \label{eqn:ncdef}
\end{eqnarray}
Here, $J^{B}_{\mu}$ is the baryon current, defined to be coupled to
$\vbg_{(s)}^{\mu}$. These relations are consistent with the charge
algebra $Q=T^{0}+B/2$ (where $B$ is the baryon number). $V^{i\mu}$ and
$A^{i\mu}$ are the isovector vector current and the  isovector
axial-vector current, respectively. We do not discuss ``seagull''
terms of higher order in the couplings because they do not enter in
our calculations \cite{EMQHD07,XZThesis}.

\section{Form factors for currents} \label{app:ff}

Here we use matrix elements of the various currents to define the
form factors produced by the Lagrangian \cite{FST97}. 
By using information presented in Appendix.~\ref{app:chisym} and the Lagrangian
in Sec.~\ref{sec:LnoDelta},  we can determine the matrix elements.
\begin{widetext}
\begin{eqnarray}
\bra{N, B} V^{i}_{\mu} \ket{N, A}&=& \left[
\psibar{u}_{f}\dgamma{\mu}u_{i}
          + \frac{\beta^{(1)}}{M^{2}}\,\psibar{u}_{f}(q^{2}\dgamma{\mu}
          -\slashed{q} q_{\mu}) u_{i} -\frac{g_{\rho}}{g_{\gamma}}
          \frac{q^{2}g_{\mn}-q_{\mu}q_{\nu}}{q^{2}-m^{2}_{\rho}} \,
          \psibar{u}_{f}\ugamma{\nu}u_{i} \right]\, \bra{B} \uhalftau{i}\ket{A} \notag
          \\[5pt]
&{}&+\left[
          2\lambda^{(1)}\,\psibar{u}_{f}\frac{\dsigma{\mn}iq^{\nu}}{2M}\, u_{i}
          -\frac{f_{\rho}g_{\rho}}{g_{\gamma}} \frac{q^{2}}{q^{2}-m^{2}_{\rho}}\,
          \psibar{u}_{f} \frac{\dsigma{\mn}iq^{\nu}}{2M}\, u_{i} \right]\,
          \bra{B} \uhalftau{i}\ket{A} \ ,  \\       
\bra{N, B} J^{B}_{\mu} \ket{N, A}&=&  \left[ \psibar{u}_{f}\dgamma{\mu}u_{i}
          + \frac{\beta^{(0)}}{M^{2}}\, \psibar{u}_{f}(q^{2}\dgamma{\mu}
          -\slashed{q} q_{\mu}) u_{i} -\frac{2 g_{v}}{3 g_{\gamma}}
          \frac{q^{2}g_{\mn}-q_{\mu}q_{\nu}}{q^{2}-m^{2}_{v} }\,
          \psibar{u}_{f}\ugamma{\nu}u_{i} \right]\, \delta_{B}^{A} \notag
          \\[5pt]
&&+\left[ 2\lambda^{(0)}\, \psibar{u}_{f}\frac{\dsigma{\mn}iq^{\nu}}{2M}\,u_{i}
          -\frac{2 f_{v}g_{v}}{3 g_{\gamma}}
          \frac{q^{2}}{q^{2}-m^{2}_{v}}\,
          \psibar{u}_{f} \frac{\dsigma{\mn}iq^{\nu}}{2M}\, u_{i} \right]\,
          \delta_{B}^{A} \ , \\
 \bra{N, B; \pi,j,k_{\pi}} A^{i}_{\mu} \ket{N, A} &=&
 -\frac{\epsilon^{i}_{\,jk}}{f_{\pi}}\, \bra {B} \uhalftau{k}\ket{A}\,
          \psibar{u}_{f}\ugamma{\nu}u_{i}\, \left[g_{\mn}
          +\frac{\beta^{(1)}}{M^{2}}\, (q\cdot(q-k_{\pi}) g_{\mn}-(q-k_{\pi})_{\mu} q_{\nu})
          \right. \notag \\[5pt]
&{}& \left. {}-\frac{g_{\rho}}{g_{\gamma}}\,\frac{q\cdot(q-k_{\pi})g_{\mn}-(q-k_{\pi})_{\mu}q_{\nu}}
          {(q-k_{\pi})^{2}-m^{2}_{\rho}} \right]   \notag \\[5pt]
&&-\frac{\epsilon^{i}_{\,jk}}{f_{\pi}}\, \bra {B} \uhalftau{k}\ket{A}\,
          \psibar{u}_{f}\frac{\dsigma{\mn}iq^{\nu}}{2M}\, u_{i}
          \left[2\lambda^{(1)}-\frac{f_{\rho}g_{\rho}}{g_{\gamma}}\,
          \frac{q\cdot(q-k_{\pi})}{(q-k_{\pi})^{2}-m^{2}_{\rho}} \right] \ . 
\end{eqnarray}
\end{widetext}

Now we consider $\bra{N, B} A^{i}_{\mu} \ket{N, A}$ and $\bra{N, B;
\pi,j} V^{i}_{\mu} \ket{N, A}$. In the chiral limit, we find
\begin{widetext}
\begin{align}
\bra{N, B} A^{i}_{\mu} \ket{N, A}=&-\bra{B} \uhalftau{i}\ket{A}
          \psibar{u}_{f} \ugamma{\nu}\ugammafive u_{i}\left[
          g_{A}\left(g_{\mn}-\frac{q_{\mu}q_{\nu}}{q^{2}}\right)
          -\frac{\beta^{(1)}_{A}}{M^{2}}(q^{2}g_{\mn}
          -q_{\mu}q_{\nu}) \right. \notag \\[5pt] 
          &\left. -2c_{a_{1}}g_{a_{1}}\,
          \frac{q^{2}g_{\mn}-q_{\mu}q_{\nu}}{q^{2}-m^{2}_{a_{1}}} \right]
          \ , \\[5pt]
 \bra{N, B; \pi,j,k_{\pi}} V^{i}_{\mu} \ket{N, A} =&\frac{\epsilon^{i}_{\,jk}}{f_{\pi}}\, \bra {B}
\uhalftau{k}\ket{A}\,
          \psibar{u}_{f}\ugamma{\nu}\ugammafive u_{i}\, \left[g_{A}g_{\mn}
          -\frac{\beta^{(1)}_{A}}{M^{2}}\, (q\cdot(q-k_{\pi})
          g_{\mn}-(q-k_{\pi})_{\mu} q_{\nu})
          \right. \notag \\[5pt]
&\left. {}-2c_{a_{1}}g_{a_{1}}
          \,\frac{q\cdot(q-k_{\pi})g_{\mn}-(q-k_{\pi})_{\mu}q_{\nu}}
          {(q-k_{\pi})^{2}-m^{2}_{a_{1}}} \right]\ .
\end{align}
\end{widetext}
Suppose that there is only one manifestly chiral-symmetry-breaking
term, i.e., the mass term for pions; then the pion-pole contribution
associated with the $g_{A}$ coupling in $\bra{N, B} A^{i}_{\mu}
\ket{N, A}$ will become
$g_{A}[g_{\mn}-q_{\mu}q_{\nu}/(q^{2}-m_{\pi}^{2})]$, while the other
parts in $\bra{N, B} A^{i}_{\mu} \ket{N, A}$, as well as the whole
$\bra{N, B; \pi,j} V^{i}_{\mu} \ket{N, A}$, will remain unchanged.
However, we must realize that there are other possible
chiral-symmetry-breaking terms contributing to $\bra{N, B}
A^{i}_{\mu} \ket{N, A}$. For example, $(m_{\pi}^{2}/M)\,
\psibar{N}i\ugammafive (U-U^{\dagger})N$ can contribute to $\bra{N,
B} A^{i}_{\mu} \ket{N, A}$ as
\begin{eqnarray}
-\frac{2m_{\pi}^{2}}{M^{2}} \frac{q_{\mu} \slashed{q}
\ugammafive}{q^{2}-m_{\pi}^{2}}\bra{B} \uhalftau{i}\ket{A}\ . \notag
\end{eqnarray}
To simplify the fitting procedures, we use the following form
factors [$G_{A}^{md}$ can be found in Eq.~(\ref{eqn:defofGA1})]:
\begin{widetext}
\begin{align}
\bra{N, B} A^{i}_{\mu} \ket{N, A}&= -G_{A}^{md}(q^{2})\bra{B}
          \uhalftau{i}\ket{A} \psibar{u}_{f} \left(g_{\mn}
          -\frac{q_{\mu}q_{\nu}}{q^{2}-m_{\pi}^{2}} \right) \ugamma{\nu}\ugammafive u_{i}
          \ , \\[5pt]
\bra{N, B; \pi,j,k_{\pi}} V^{i}_{\mu} \ket{N,
          A}&=\frac{\epsilon^{i}_{\,jk}}{f_{\pi}}\, \bra {B}
          \uhalftau{k}\ket{A}\,
          \psibar{u}_{f}\ugamma{\nu}\ugammafive u_{i}\,
          \left[g_{A}g_{\mn} +\delta G^{md}_{A}[(q-k_{\pi})^{2}] \,
          \frac{q\cdot(q-k_{\pi})g_{\mn}-(q-k_{\pi})_{\mu}q_{\nu}}
          {(q-k_{\pi})^{2}} \right] \ .
\end{align}
\end{widetext}
Finally, we calculate the pion form factor $\bra{\pi,k}
V^{i}_{\mu}\ket{\pi,j}$:
\begin{widetext}
\begin{align}
\bra{\pi,k,k_{\pi}} V^{i}_{\mu}\ket{\pi,j,k_{\pi}-q} &=i\epsilon^{ij}_{\;\; k}
          (2k_{\pi}-q)_{\mu} +2i\, \frac{g_{\rho\pi\pi}}{g_{\gamma}}\,  \epsilon^{ij}_{\;\;k}\,
          \frac{q^{2}}{m^{2}_{\rho}} \frac{1}{q^{2}-m^{2}_{\rho}}\,
          (q\cdot k_{\pi} q_{\mu}-q^{2}k_{\pi\mu}) \notag \\[5pt]
q^{2}\to m_{\rho}^{2} \ \text{in numerator} &\longrightarrow 
          i\epsilon^{ij}_{\;\; k}\, (2k_{\pi}-q)_{\mu} +2i\, \frac{g_{\rho\pi\pi}}{g_{\gamma}}\, \epsilon^{ij}_{\;\;k}\,
          \frac{1}{q^{2}-m^{2}_{\rho}}\, (q\cdot k_{\pi} q_{\mu}-q^{2}k_{\pi\mu}) 
          \ . 
\end{align}
\end{widetext}

\section{power counting for diagrams with $\Delta$} \label{app:deltapowercounting}
Including $\Delta$ resonances in calculations, we have a new mass scale $\delta \equiv
m-M \approx 300$ MeV. We must also consider the order of the
$\Delta$ width $\Gamma$. Formally, it is counted as $
O(Q^{3}/M^{2})$; however, numerical calculations with
Eq.~(\ref{eqn:tang98width}) indicate that it should be counted as
$O(10 Q^{3}/M^{2})$. Because of these two issues, we have to
rethink the power counting of diagrams involving $\delta$ in two
energy regimes. One is near the resonance, while the other is at
lower energies, away from the resonance. In the resonance region,
the $\Delta$ propagator scales like
\begin{align}
S_{F} &\sim \frac{1}{i \Gamma} + O \left(\frac{1}{M} \right) \approx
\frac{1}{10 i\, O(Q^{3}/M^{2})} \approx \frac{1}{i\, O(Q^{2}/M)} \notag \\[5pt]
&\sim \frac{1}{O(Q)} \frac{M}{i\, O(Q)}\ ,
\end{align}
where the $O({1}/{M})$ comes from non-pole terms. In the
lower-energy region,
\begin{align}
S_{F} &\sim  \frac{1}{2[\delta-O(Q)] - 10 i\, O(Q^{3}/M^{2})} +O
\left(\frac{1}{M} \right) \notag \\[5pt]
& \sim \frac{1}{O(Q)} \frac{O(Q)}{2 \delta} +O
\left(\frac{1}{M} \right)\approx \frac{1}{O(Q)} \frac{O(Q)}{M}\ .
\end{align}
So compared to the normal power counting mentioned above, in which
the nucleon propagator scales as ${1}/{O(Q)}$, for diagrams
involving one $\Delta$ in the $s$ channel, we take $\nu \to \nu -1$
in the resonance regime and $\nu \to \nu+1 $ away from the
resonance.

\section{Renormalized $\Delta$ propagator} \label{app:deltaprop}
In this work, $\Delta$'s propagator \cite{tang98} is dressed as 
\begin{widetext}
\begin{eqnarray}
S_{F}^{\mn}(p) &\equiv&
          -\frac{\slashed{p}+m}{p^{2}-m^{2}-\Pi(p^{2})+im\Gamma(p^{2})}
          P^{(\frac{3}{2}) \mn}-\frac{1}{\sqrt{3}m}P^{(\half)
          \mn}_{12}-\frac{1}{\sqrt{3}m}P^{(\half) \mn}_{21} 
          +\frac{2}{3m^{2}}\,(\slashed{p}+m) P^{(\half) \mn}_{22}
          \notag \\[5pt]
&&{} + O(\Gamma /m) \times \text{non-pole terms,}  \\[5pt]
\Gamma(p^{2}) &=& \frac{\pi}{12mp^{4}}\, \frac{h_{A}^{2}}{(4\pi
          f_{\pi})^{2}}\, (p^{2}+M^{2}+2Mm)  \times\left[(p^{2}-M^{2})^{2}-(p^{2}+3M^{2})m_{\pi}^{2}\right]
          \sqrt{(p^{2}-M^{2})^{2}-4p^{2}m_{\pi}^{2}}\ .
          \label{eqn:tang98width} 
\end{eqnarray}
\end{widetext}
Here:
\begin{align}
P^{(\frac{3}{2}) \mn}&= g^{\mn}-\frac{1}{3}\, \ugamma{\mu}\ugamma{\nu}
          +\frac{1}{3p^{2}}\, \ugamma{[\mu}p^{\nu]} \slashed{p}
          -\frac{2}{3p^{2}}\, p^{\mu}p^{\nu} \ , 
          \\[5pt]
P^{(\frac{1}{2}) \mn}_{11}&=\frac{1}{3}\, \ugamma{\mu}\ugamma{\nu}
          -\frac{1}{3p^{2}}\, \ugamma{[\mu} p^{\nu]} \slashed{p}
          -\frac{1}{3p^{2}}\, p^{\mu}p^{\nu} \ , \\[5pt]
P^{(\frac{1}{2}) \mn}_{12}&= \frac{1}{\sqrt{3}p^{2}}\,
          (-p^{\mu}p^{\nu}+\ugamma{\mu}p^{\nu}\slashed{p}) \ , \\[5pt]
P^{(\frac{1}{2}) \mn}_{21}&=-P^{(\frac{1}{2}) \nu\mu}_{12}\ , \\[5pt] 
P^{(\frac{1}{2}) \mn}_{22}&=\frac{1}{p^{2}}\, p^{\mu} p^{\nu} \ .
\end{align}
We take $m=1232$ MeV as the Breit--Wigner mass \cite{pdg2008} and set $\Pi=0$. 
Note that $\Gamma$ is implicitly associated with a factor of
$\Theta[p^{2}-(M+m_{\pi})^{2}]$. And no singularity exists in this propagator 
at $p^{2}=0$.

\section{kinematics} \label{app:kinmatics}

Following a standard calculation, we find the total cross section:
\begin{widetext}
\begin{eqnarray}
\sigma &=& \int \frac{\overline{\vert M \vert^{2}}}{4
          \vert p_{li}^{L} \cdot p_{ni}^{L}\vert}\, (2\pi)^{4}
          \delta^{(4)}\left(\sum_{i}p_{i}^{L}\right)  \frac{d^{3}
          \vec{p}_{lf}^{\mkern3mu L}}{(2\pi)^{3}2E_{lf}^{L}}
          \frac{d^{3} \vec{p}_{\pi}^{\mkern3mu L}}{(2\pi)^{3}2E_{\pi}^{L}}
          \frac{d^{3} \vec{p}_{nf}^{\mkern3mu L}}{(2\pi)^{3}2E_{nf}^{L}} \notag
          \\[5pt]
&=&\int \frac{\overline{\vert M \vert^{2}}}
          {4 \vert p_{li}^{L} \cdot p_{ni}^{L}\vert}\, (2\pi)^{4}
          \delta(q^{0}+p_{ni}^{0}-p_{nf}^{0}-p_{\pi}^{0})\,
          \frac{1}{(2\pi)^{3}2E_{nf}}
          \frac{d^{3} \vec{p}_{lf}^{\mkern3mu L}}{(2\pi)^{3}2E_{lf}^{L}}
          \frac{d^{3} \vec{p}_{\pi}}{(2\pi)^{3}2E_{\pi}} \notag
          \\[5pt]
&=&\int \frac{\overline{\vert M \vert^{2}}}{32 M_{n}}
          \frac{1}{(2\pi)^{5}}  \frac{\modular{p}{\pi}}{E_{\pi}+E_{nf}}
          \frac{\vert \vec{p}_{lf}^{\mkern3mu L}\vert }
          {\vert \vec{p}_{li}^{\mkern3mu L}\vert}\,
          d\Omega_{\pi}\, dE_{lf}^{L}\, d\Omega_{lf}^{L} \ .  
\end{eqnarray}
\end{widetext}

The variables without and ``$L$'' superscript are measured in the isobaric frame (where $\Delta$ is static). 
It is quite complicated to calculate the boundary of phase space in
terms of the integration variables in the preceding equations.
Later, we will work out the boundary of phase space in terms of the
invariant variables $Q^{2}$ and $M_{\pi n}$ in the c.m. frame of the
whole system, so we would like to have the following:
\begin{align}
Q^{2}&= -M_{lf}^{2}+2E_{li}^{L}(E_{lf}^{L}-\vert
          \vec{p}_{lf}^{\mkern3mu L}\vert \cos{\theta_{lf}^{L}} )\ , 
          \\[5pt]
M_{\pi n}^{2}&=(q^{L}+p_{ni}^{L})^{2}=-Q^{2}+M_{n}^{2}+2M_{n}(E_{li}^{L}-E_{lf}^{L})
          \ , \\[5pt] 
& dQ^{2} dM_{\pi n}^{2} =  4M_{n} E_{li}^{L}  \, \vert
          \vec{p}_{lf}^{\mkern3mu L}\vert dE_{lf}^{L}\, d\cos{\theta_{lf}^{L}}\ . 
\end{align}
By using the invariance of the cross section with respect to
rotations around the incoming lepton direction, we have $\int
d\Omega_{lf}^{L}=\int d\cos{\theta_{lf}^{L}\, 2\pi}$, and thus
\begin{align}
\sigma=\int \frac{\overline{\vert M \vert^{2}}}{64 M_{n}^{2}}
          \frac{1}{(2\pi)^{5}}  \frac{\modular{p}{\pi}}{E_{\pi}+E_{nf}}
          \frac{\pi}{\vert \vec{p}_{li}^{\mkern3mu L}\vert E_{li}^{L}}\,
          d\Omega_{\pi}\,
          dM_{\pi n}^{2}\, d Q^{2} \ . 
\end{align}

In the isobaric frame, there is no constraint on the direction of
the outgoing pion due to the kinematics. Thus the boundary of $ \Omega_{\pi}$ is the whole
solid angle in the isobaric frame. Now let's work out the boundary
of phase space in the c.m. frame. We have
\begin{align}
M_{A}^{2}&\equiv p_{A}^{2}=(p^{L}_{ni}+p^{L}_{li})^{2}
          =(M_{n}+E^{L}_{li})^{2}-(E^{L}_{li})^{2} \notag \\[5pt] 
         &=M_{n}^{2}+2M_{n}E^{L}_{li} \ , \\[5pt]
M_{\pi n}^{2}&\equiv (p_{\pi}+p_{nf})^{2}
          =(p^{C}_{A}-p^{C}_{lf})^{2} \notag \\[5pt]  
         &=M_{A}^{2}+M_{lf}^{2}-2M_{A}E^{C}_{lf}\ .
\label{eqn:MpinElf}
\end{align}
Here, $E^{C}_{lf}$ is the final lepton's energy in the c.m. frame.
From now on, all the quantities in the c.m. frame will be labeled in this
way. So, for given $E^{L}_{li}$, i.e., $M_{A}$, we can see that
\begin{eqnarray}
M_{n}+M_{\pi} \leqslant M_{\pi n} \leqslant M_{A}-M_{lf}  \ . 
\end{eqnarray}
By using Eq.~(\ref{eqn:MpinElf}), we find
\begin{eqnarray}
(E^{C}_{lf})_{ \mathrm{max(min)}}=\frac{M_{A}^{2}+M_{lf}^{2}
          -(M_{\pi n }^{2})_{\mathrm{min(max)}}}{2M_{A}}\ . 
\end{eqnarray}
Then, for given $E^{L}_{li}$ and $M_{\pi n}$ (or $ E^{C}_{lf})$,
using
$Q^{2}=-M_{lf}^{2}+2E^{C}_{li}E^{C}_{lf}-2E^{C}_{li}\modular{p}{lf}^{C}
\cos{\theta^{C}_{lf}}$ [where $\theta^{C}_{lf}$ is the angle between
the outgoing lepton's direction and the incoming lepton's direction
in the c.m. frame, and $E_{li}^{C}=(M_{A}^{2}-M_{n}^{2})/2M_{A}$ is
the initial lepton's energy in the c.m. frame], we finally arrive at
\begin{align}
[Q^{2}(E^{C}_{lf})]_{\mathrm{min}} &=
          -M_{lf}^{2}+\frac{2E^{C}_{li}M_{lf}^{2}}
          {E^{C}_{lf}+\sqrt{(E^{C}_{lf})^{2}-M_{lf}^{2}}} \ , 
          \\[5pt]
[Q^{2}(E^{C}_{lf})]_{\mathrm{max}} &= -M_{lf}^{2}+ 2E^{C}_{li}
\left(
          E^{C}_{lf}+\sqrt{(E^{C}_{lf})^{2}-M_{lf}^{2}}\, \right) \ . 
\end{align}
These equations give a description of the phase-space boundary in
terms of the invariants $M_{\pi n}$ and $Q^{2}$.

\end{document}